\documentclass[12pt,twoside]{article}
\usepackage[latin1]{inputenc}
\usepackage{times}
\usepackage{epsfig}
\usepackage{amsfonts}
\usepackage{float}
\usepackage{amsmath}
\usepackage{latexsym}
\usepackage{color}
\usepackage[square,sort,comma,numbers]{natbib}
\usepackage{bm}

\usepackage{xr} 

\usepackage{times,epsfig,amsmath,amssymb,graphicx,latexsym,amsthm}
\usepackage[latin1]{inputenc}
\usepackage{wrapfig,lipsum,booktabs}
\usepackage[mathscr]{eucal}
\usepackage{amsfonts}
\usepackage{layout}
\usepackage{color}
\usepackage{bm}
\usepackage{eurosym}
\usepackage{titlesec}
\usepackage{cutwin}
\usepackage[normalem]{ulem}
\usepackage{amsmath}
\usepackage{arydshln}
\usepackage{multirow}
\usepackage{longtable}

\relax 
\allowdisplaybreaks[3]

\relax

\newcommand{\bay}{\begin{array}}
\newcommand{\eay}{\end{array}}

\newcommand{\bry}{\begin{array}}
\newcommand{\ery}{\end{array}}

\newcommand{\bqa}{\begin{eqnarray*}}
\newcommand{\eqa}{\end{eqnarray*}}

\newcommand{\bqan}{\begin{eqnarray}}
\newcommand{\eqan}{\end{eqnarray}}

\newcommand{\bqt}{\begin{quote}}
\newcommand{\eqt}{\end{quote}}

\newcommand{\bt}{\begin{tabbing}}
\newcommand{\et}{\end{tabbing}}

\newcommand{\bit}{\begin{itemize}}
\newcommand{\eit}{\end{itemize}}

\newcommand{\ben}{\begin{enumerate}}
\newcommand{\een}{\end{enumerate}}

\newcommand{\beq}{\begin{equation}}
\newcommand{\eeq}{\end{equation}}

\newcommand{\bdes}{\begin{description}}
\newcommand{\edes}{\end{description}}

\newcommand{\btb}{\begin{tabular}}
\newcommand{\etb}{\end{tabular}}

\newcommand{\bpic}{\begin{picture}}
\newcommand{\epic}{\end{picture}}

\newcommand{\bcen}{\begin{center}}
\newcommand{\ecen}{\end{center}}

\newcommand{\bfg}{\begin{figure}}
\newcommand{\efg}{\end{figure}}

\newcommand{\bmp}{\begin{minipage}}
\newcommand{\emp}{\end{minipage}}

\newcommand{\bgan}{\begin{gather}}
\newcommand{\egan}{\end{gather}}

\newcommand{\bal}{\begin{align*}}
\newcommand{\eal}{\end{align*}}

\newcommand{\baln}{\begin{align}}
\newcommand{\ealn}{\end{align}}

\newcommand{\bala}{\begin{alignat*}}
\newcommand{\eala}{\end{alignat*}}

\newcommand{\balan}{\begin{alignat}}
\newcommand{\ealan}{\end{alignat}}

\newcommand{\bspt}{\begin{split}}
\newcommand{\espt}{\end{split}}

\newcommand{\vepsilon}{\vep}

\newtheorem{definition}{{\sc Definition}\sc}[section]
\newcommand{\bdefi}{\begin{definition}}
\newcommand{\edefi}{\end{definition}}

\newtheorem{appropr}[definition]{{\sc Approximation Procedure}\sc}
\newcommand{\bappr}{\begin{appropr}}
\newcommand{\eappr}{\end{appropr}}

\newtheorem{bedi}[definition]{{\sc Condition}\sc}
\newcommand{\bbd}{\begin{bedi}}
\newcommand{\ebd}{\end{bedi}}

\newtheorem{bedin}[definition]{{\sc Conditions}\sc}
\newcommand{\bbdn}{\begin{bedin}}
\newcommand{\ebdn}{\end{bedin}}

\newtheorem{corollary}[definition]{{\sc Corollary}\sc}
\newcommand{\bco}{\begin{corollary}}
\newcommand{\eco}{\end{corollary}}

\newtheorem{lemma}[definition]{{\sc Lemma}\sc}
\newcommand{\blem}{\begin{lemma}}
\newcommand{\elem}{\end{lemma}}

\newtheorem{proposition}[definition]{{\sc Proposition}\sc}
\newcommand{\bpro}{\begin{proposition}}
\newcommand{\epro}{\end{proposition}}

\newtheorem{satz}[definition]{{\sc Theorem}\sc}
\newcommand{\bsa}{\begin{satz}}
\newcommand{\esa}{\end{satz}}

\newtheorem{assumption}[definition]{{\sc Assumption}\sc}
\newcommand{\bas}{\begin{assumption}}
\newcommand{\eas}{\end{assumption}}

\newtheorem{assumptions}[definition]{{\sc Assumptions}\sc}
\newcommand{\bass}{\begin{assumptions}}
\newcommand{\eass}{\end{assumptions}}

\newtheorem{abb}{{\sc Figure}\sc}[section]
\newcommand{\babb}{\begin{abb}}
\newcommand{\eabb}{\end{abb}}

\newenvironment{remark}{\begin{rmk}\sl}{\end{rmk}}
\newtheorem{rmk}{{\sc Remark}\sc}[section]
\newcommand{\brem}{\begin{remark}}
\newcommand{\erem}{\end{remark}}

\newenvironment{example}{\begin{exmp}\rm}{\end{exmp}}
\newtheorem{exmp}{{\sc Example}\sc}[section]
\newcommand{\bbsp}{\begin{example}}
\newcommand{\ebsp}{\end{example}}
\newcommand{\bexa}{\begin{example}}
\newcommand{\eexa}{\end{example}}

\newtheorem{model}{{\sc Model}\sc}[section]
\newcommand{\bmdl}{\begin{model}}
\newcommand{\emdl}{\end{model}}

\newtheorem{scheme}{{\sc Scheme}\sc}[section]
\newcommand{\bscm}{\begin{scheme}}
\newcommand{\escm}{\end{scheme}}

\newenvironment{tabelle}{\begin{tabl}\sl}{\end{tabl}}
\newtheorem{tabl}{{\sc Table}\sc}[section]
\newcommand{\btab}{\begin{tabelle}}
\newcommand{\etab}{\end{tabelle}}

\newenvironment{exercise}{\begin{exc}}{\end{exc}}
\newtheorem{exc}{{\sc Exercise}\sc}
\newcommand{\bexe}{\begin{exercise}}
\newcommand{\eexe}{\end{exercise}}

\newcommand{\diag}{\operatorname{\it diag}}

\newcommand{\olM}{\overline{M}}

\newcommand{\olX}{\overline{X}}
\newcommand{\olY}{\overline{Y}}

\newcommand{\va}{\boldsymbol{a}}
\newcommand{\vb}{\boldsymbol{b}}
\newcommand{\vc}{\boldsymbol{c}}

\newcommand{\ve}{\boldsymbol{e}}

\newcommand{\vp}{\boldsymbol{p}}

\newcommand{\vA}{\boldsymbol{A}}
\newcommand{\vB}{\boldsymbol{B}}

\newcommand{\vD}{\boldsymbol{D}}
\newcommand{\vE}{\boldsymbol{E}}

\newcommand{\vH}{\boldsymbol{H}}
\newcommand{\vI}{\boldsymbol{I}}

\newcommand{\vM}{\boldsymbol{M}}

\newcommand{\vQ}{\boldsymbol{Q}}

\newcommand{\vX}{\boldsymbol{X}}
\newcommand{\vY}{\boldsymbol{Y}}

\newcommand{\vbeta}{\boldsymbol{\beta}}

\newcommand{\vGamma}{\boldsymbol{\Gamma}}

\newcommand{\vep}{\boldsymbol{\epsilon}}

\newcommand{\vSigma}{\boldsymbol{\Sigma}}

\newcommand{\vwhb}{\boldsymbol{\widehat{b}}}

\newcommand{\vwhp}{\boldsymbol{\widehat{p}}}

\newcommand{\vwhGamma}{\boldsymbol{\widehat{\Gamma}}}

\newcommand{\vwhSigma}{\boldsymbol{\widehat{\Sigma}}}

\newcommand{\vwtp}{\boldsymbol{\widetilde{p}}}

\newcommand{\vwtX}{\boldsymbol{\widetilde{X}}}

\newcommand{\whb}{\widehat{b}}

\newcommand{\whp}{\widehat{p}}

\newcommand{\whdelta}{\widehat{\delta}}

\newcommand{\whsigma}{\widehat{\sigma}}

\newcommand{\wtn}{\widetilde{n}}

\newcommand{\wtX}{\widetilde{X}}

\begin{document}


\title{\Large \bf The Behrens-Fisher Problem with Covariates and Baseline Adjustments}
\author{Cong Cao$^*$, Markus Pauly$^{**}$ and Frank Konietschke$^{*}$ 
}
\maketitle

\begin{abstract}
The Welch-Satterthwaite $t$-test is one of the most prominent and often used statistical inference method in applications. The method is, however, not flexible with respect to adjustments for baseline values or other covariates, which may impact the response variable. Existing analysis of covariance methods are typically based on the assumption of equal variances across the groups. This assumption is hard to justify in real data applications and the methods tend to not control the type-1 error rate satisfactorily under variance heteroscedasticity. In the present paper, we tackle this problem and develop unbiased variance estimators of group specific variances, and especially of the variance of the estimated adjusted treatment effect in a general analysis of covariance model. These results are used to generalize the Welch-Satterthwaite $t$-test to covariates adjustments. Extensive simulation studies show that the method accurately controls the nominal type-1 error rate, even for very small sample sizes, moderately skewed distributions and under variance heteroscedasticity. A real data set motivates and illustrates the application of the proposed methods. 
\end{abstract}


\vfill
\vfill

\noindent${}^{*}$ {Department of Mathematical Sciences, The University of Texas at Dallas, 800 W Campbell Road, 75080 Richardson, TX, USA\\
 \mbox{ }\hspace{1 ex}email: fxk141230@utdallas.edu}

\noindent${}^{**}$ {Institute of Statistics, Ulm University, Helmholtzstr. 20, 89081 Ulm, Germany}
\newpage

 \newpage

\section{Introduction} 
\noindent The statistical comparison of two independent samples is naturally arising in a variety of different disciplines, e.g., in biological, ecological, psychological, or medical studies. When data is measured on a metric scale, roughly symmetrically distributed and assumed to have equal variances (homogeneous), the $t$-test is often used for making inferences in the means $\mu_1$ and $\mu_2$ of the two distributions. In case of unequal variances, the Welch $t$-test 
\begin{eqnarray}\label{ttest}
T = \frac{\overline{X}_{1\cdot} - \overline{X}_{2\cdot}-(\mu_1-\mu_2)}{\sqrt{\frac{s_1^2}{n_1} + \frac{s_2^2}{n_2}}} 
\end{eqnarray}
is typically applied \cite{Satterthwaite1946df, Welch1947ttest}. Here, $\overline{X}_{i\cdot} = n_i^{-1} \sum_{k=1}^{n_i} X_{ik}$ and $s_i^2 = (n_i-1)^{-1}\sum\limits_{k=1}^{n_i} (X_{ik} - \overline{X}_{i\cdot})^2$ denote the empirical means and variances of the independent random samples $X_{i1},\ldots,X_{in_i}$ coming from distribution $F_i, i=1,2$, respectively. Under the assumption of normality of the data, $X_{ik} \sim N(\mu_i,\sigma_i^2), k=1,\ldots,n_i$, the distribution of $T$ in (\ref{ttest}) can be approximated by a $t_\nu$-distribution, where the degree of freedom
\begin{eqnarray} \label{nu}
\nu = \frac{\left(\frac{s_1^2}{n_1} + \frac{s_2^2}{n_2} \right)^2}{\frac{s_1^4}{n_1^2(n_1-1)} + \frac{s_2^4}{n_2^2/(n_2-1)}}
\end{eqnarray}
is known as \textit{Welch-Satterthwaite degree of freedom} (\cite{Imbenssmallsp}). It is derived by equating both the expectations and variances of the weighted sum of the sample variances $\frac{s_1^2}{n_1} + \frac{s_2^2}{n_2}$ by a scaled $g\cdot\chi_f^2$ distribution\textemdash also known as Box-type approximation in the literature (see, e.g., \cite{Patnaik1949, Box1954, Brunner1997boxtype}). The knowledge of the distributions of the sample variances $s_i^2$ is substantial in the approximation procedure, because the moments are equated with the moments of the respective $\chi^2$-distributions of the sample variances. Note that even when the assumption of normality is violated, the Welch-Satterthwaite $t$-test $T$ given in (\ref{ttest}) is still asymptotically valid for testing $H_0:\mu_1=\mu_2$ in the so-called \textit{Behrens-Fisher situation}, because $\nu \to \infty$ which implies that $T \stackrel{\mathcal{D}}{\to} N(0,1)$ as $\min\{n_1,n_2\} \to \infty$ (see, e.g., \cite{ramsey1980exacttype1, Ruxton2006tmann, kesselman2008robust, Derrick2016why}). For small samples, the quality of the approximation depends on the skewness (shapes) and the amount of variance heteroscedasticity (see, e.g., \cite{olejnik1984parametric, bathke2003nonparametric, Harwell2003summary}). Statistical methods which do not rely on the assumption of equal variances are especially meaningful when the distribution of a statistic under the alternative hypothesis is important, e.g. for the computation of confidence intervals for the effects of interest. In particular, different variances may also occur due to covariates impacting the response, for example when the outcome depends on baseline values, age, body weights, etc. The EMA guideline on adjustment for baseline covariates in clinical trials particularly states ''\textit{Baseline covariates impact the outcome in many clinical trials. Although baseline adjustment is not always necessary, in case of a strong or moderate association between a baseline covariate(s) and the primary outcome measure, adjustment for such covariate(s) generally improves the efficiency of the analysis and avoids conditional bias from chance covariate imbalance''}\cite{EMA2015}.

\noindent In such a situation, data is typically modeled by an \textit{Analysis of Covariance} (ANCOVA) model
\begin{eqnarray}\label{model}
\underbrace{\bm{Y}}_{\text{Response}} = \underbrace{\bm{Xb}}_{\text{Fixed Effects}} + \underbrace{\bm{Mp}}_{\text{Regression}} + \underbrace{\bm{\epsilon}}_{\text{Error}},
\end{eqnarray}
where $\bm{X}$ is a fixed and known design matrix, $\bm{b} = (b_1,b_2)'$ denotes the vector of fixed treatment effects (treatment/control), $\bm{M}$ denotes a matrix (full-rank or non-full-rank) of $L$ fixed covariates, $\bm{p} = (p_1,\ldots,p_L)'$ the vector of regression coefficients, and $\bm{\epsilon}$ denotes the error term  \citep{fisher1927crop}. Thus, the fixed expected location values are $b_1$ and $b_2$ in model (\ref{model}). The current gold standard for testing the hypothesis $H_0: b_1=b_2$ is to perform a classical ANCOVA $F$-test with covariates or, in the situation considered here, its two-sample $t$-test type version \textemdash which is only valid when the data have equal variances, see, e.g., the excellent textbook by \cite{searle1987linear} and references therein. In many experiments, however, data distributions cannot be modeled by a normal distribution and/or homogeneous variances, e.g., when reaction times or count data are observed. In particular, when the model assumptions are not met, the ANCOVA tends to provide rather conservative or liberal test decisions, depending on the shapes of the distributions, sample size allocations and/or degree of variance heteroscedasticity (see the extensive simulation results presented in Section~\ref{sec: simus}). Thus, there is a need for heteroscedastic ANCOVA methods and especially for a generalization of the Welch-Satterthwaite $t$-test to such scenarios.\\

\noindent The arising problem is the unbiased estimation of the variances or the covariance matrix of the estimated treatment effects $b_1$ and $b_2$ along with the computation of the degrees of freedom of its approximate $t$-distribution. Several \textit{Heteroscedasticity Consistent Standard Error} (HCSE) estimators of their covariance matrix have been developed, however, most of them are substantially biased when sample sizes are rather small \citep{hinkley1977, white1980hetero, efron1982jackknife, MacKinnon1985hetero, longervin2000, hayesli2007}. Furthermore, their sampling distributions are unknown and therefore a Box-type approximation procedure will be\textemdash if even possible\textemdash hard to compute. In the present paper, we develop unbiased estimators of the variances as well as their covariance matrix. Furthermore, we compute their sampling distributions and generalize the Welch-Satterthwaite $t$-test. It turns out that the new test can be easily computed and the degree of freedom of its remaining approximate $t$-distribution can be computed in a similar way to $\nu$ in (\ref{nu})\textemdash the variances $s_i^2$ and sample sizes $n_i$ are just replaced by the new variance estimators and weights $n_i^\ast$, which are linear combinations of the values of the covariates.  The new test procedure will be compared with the classical ANCOVA $t$-test, and a robust Wild-Bootstrap procedure for variance heteroscedastic ANCOVA models recently proposed by \cite{Zimmermann2017Wild} in extensive simulation studies. It turns out that both the adjusted Welch-Satterthwaite $t$-test and the Wild-Bootstrap method control the type-1 error rate very satisfactorily and that the methods have comparable powers to detect the alternative $H_1^{\vb}: b_1\not = b_2$. Testing for the impact of the covariates in terms of the regression parameters, the newly developed method seems to be slightly more accurate. However, the $t$-test type statistics are way less numerically intensive than the Wild-bootstrap method. In particular, their computational efficiency may play an important role in the \textit{big data} context, e.g. in genetics. Most importantly, the results obtained in the present paper allow group specific comparisons of the data by not only displaying point estimators of $b_1$ and $b_2$, but also by their group specific variances. This is highly beneficial, because they reflect the amount of variance that is explained by the regression on a group specific level.\\


\noindent The remainder of the paper is organized as follows: In Section~\ref{sec::Example} an illustrative motivating example is introduced. The statistical model, hypotheses and point estimators are discussed in Section~\ref{sec:: Model}. Unbiased estimators of the variances are derived in Section~\ref{sec:: variances}. These results will be used in Section~\ref{sec:: tests} for the derivation of the adjusted Welch-Satterthwaite $t$-test. Extensive simulation studies are presented and discussed in Section~\ref{sec: simus}. The real data set will be analyzed with the new methods in Section~\ref{sec:: evaluation} and the paper closes with a discussion about the results and future research in Section~\ref{sec:: discussion}.  All proofs are given in Appendix. \\

\noindent Throughout the manuscript the following notation will be used: Matrices are displayed in boldface. The direct sum of the matrices $\vA$ and $\vB$ is denoted by $\bm{A} \oplus \bm{B}$. Furthermore, the rank and trace of a matrix $\vA$ are given by $r(\vA)$ and $tr(\vA)$, respectively.

\section{Motivating Example}\label{sec::Example}

\noindent As a motivating example, we consider a part of the short-term study on bodyweight changes in male HSD rats being treated with specular hematite obtained from the National Toxicological Program (NTP) study number C20536. Here, we only consider the bodyweight data of the rats at week 1 (baseline) and after four weeks of treatment. Since several rats shared the same cage, we use the maximum bodyweight value per cage as the actual response value. In order to convert the data into a two-sample problem, we assign all the bodyweight values from the different dose groups to the active treatment group and keep the vehicle treated rats in the vehicle control group. In total, the values of $N=52$ rats were used, where $n_1=13$ rats were assigned to the vehicle control group and $n_2=39$ rats to the active treatment group.

\begin{figure}[h]
	\centering
		\includegraphics[height =  40 ex , keepaspectratio,angle = 0]{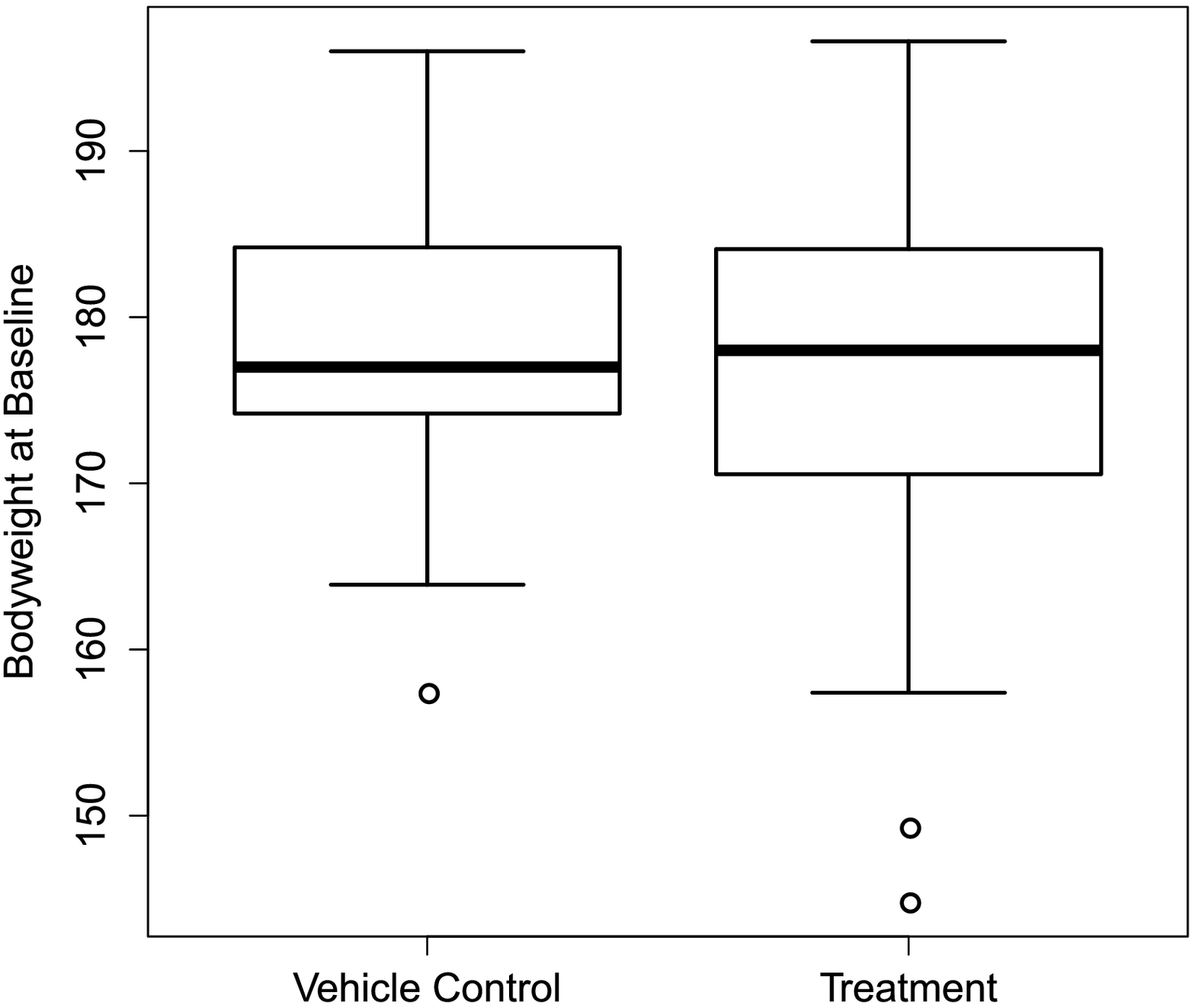}				
		\includegraphics[height =  40 ex , keepaspectratio,angle = 0]{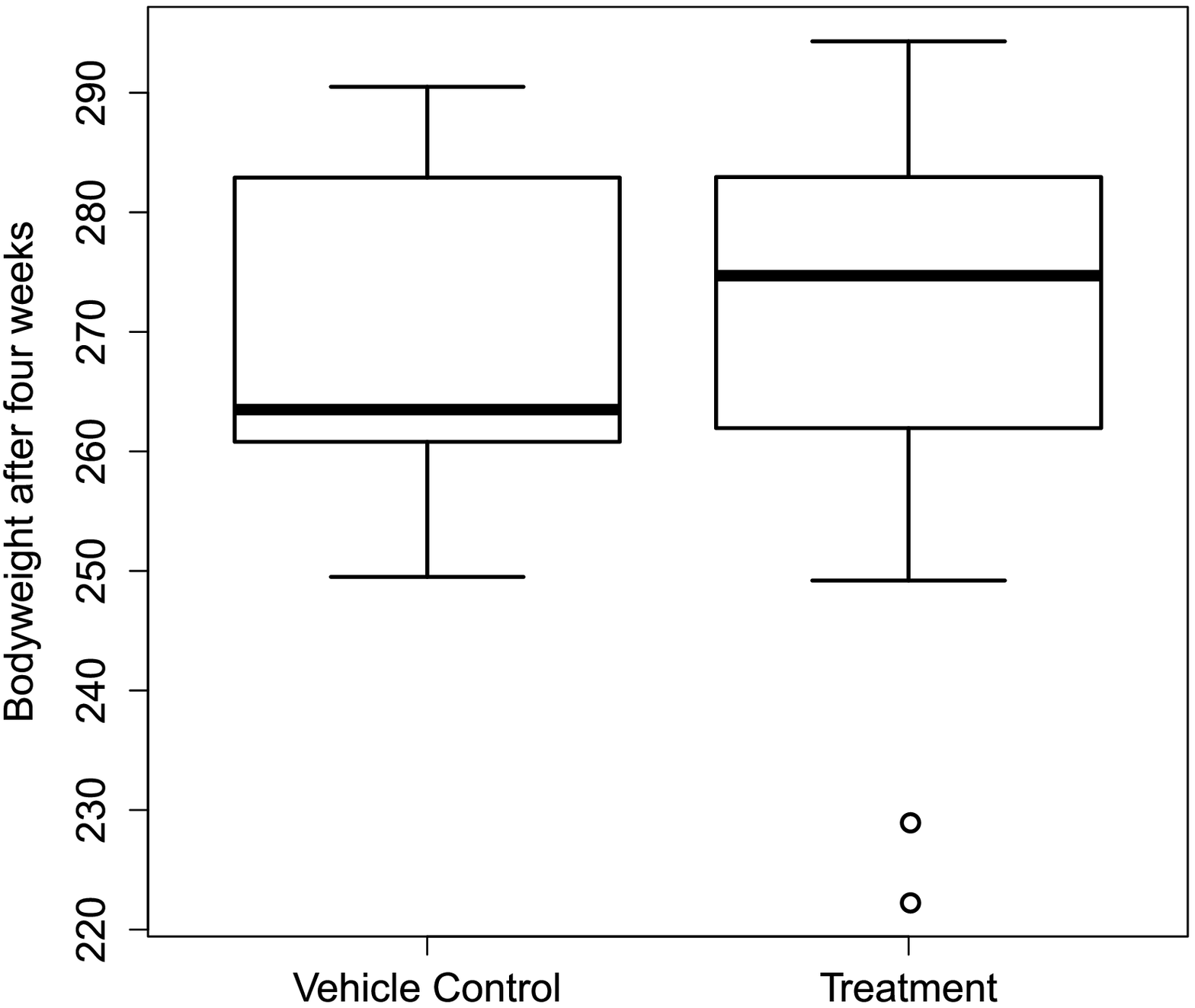}
		\caption{Boxplots of the bodyweight data at baseline (left) and after four weeks (right). }\label{Fig: boxplots}
\end{figure}

The data are displayed in Tables~\ref{dat: bodyweight0} and ~\ref{dat: bodyweight1}.
In Figure~\ref{Fig: boxplots} boxplots of the bodyweights at baseline (left) and after four weeks of treatment are displayed.

\begin{figure}[h]
	\centering
		\includegraphics[height =  25 ex , keepaspectratio,angle = 0]{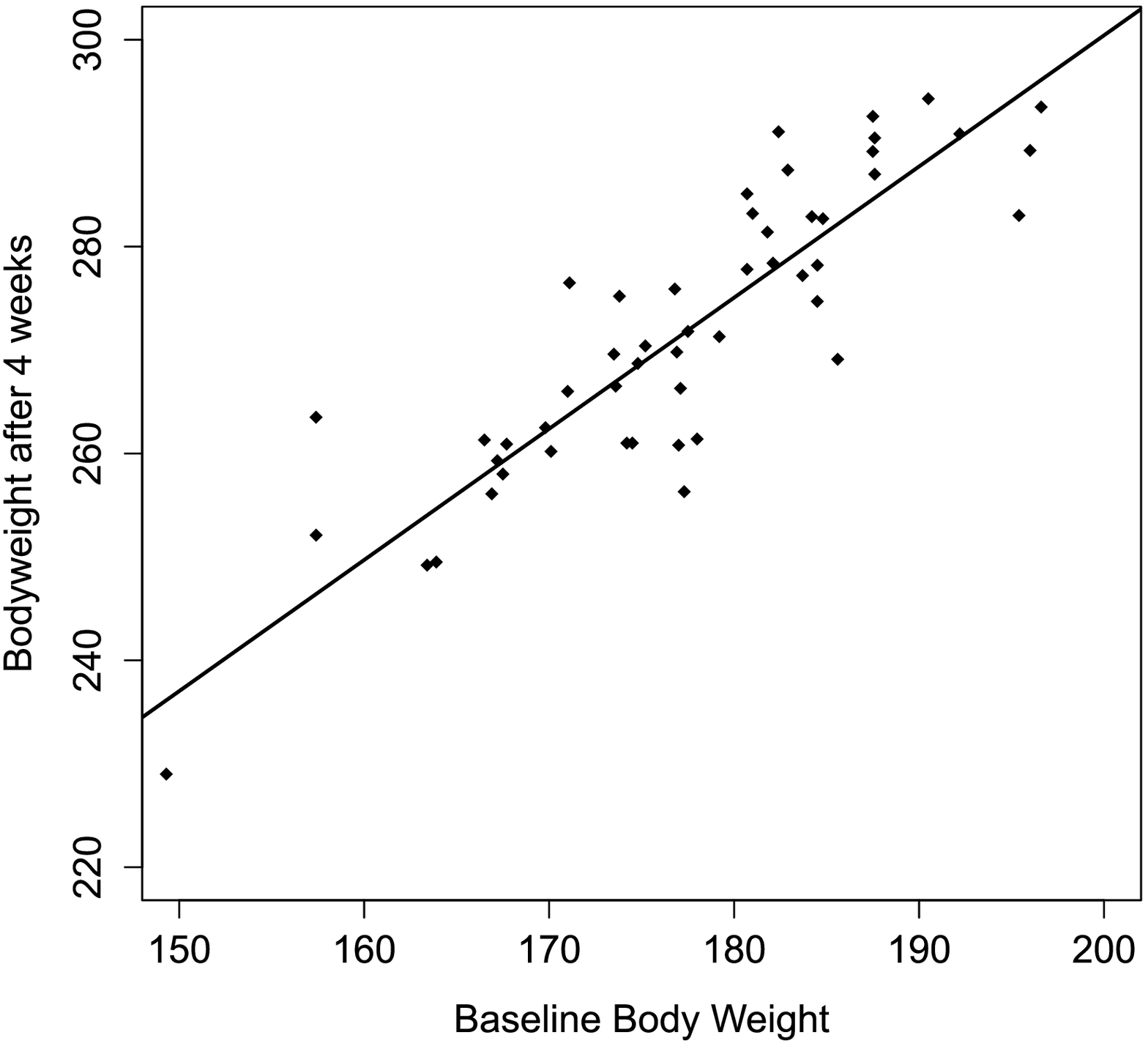} \hfill
				\includegraphics[height =  25 ex , keepaspectratio,angle = 0]{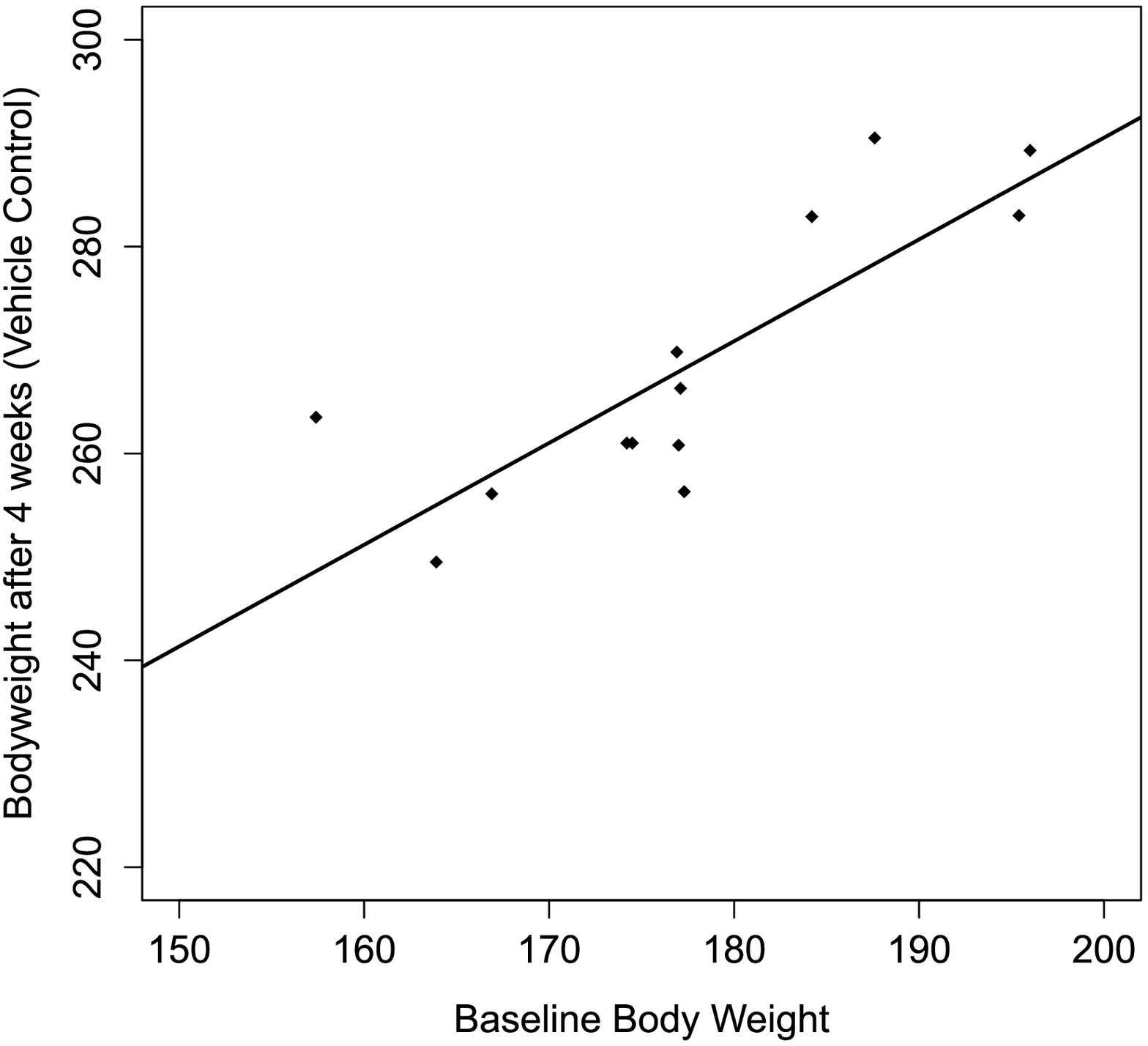} \hfill
		\includegraphics[height =  25 ex , keepaspectratio,angle = 0]{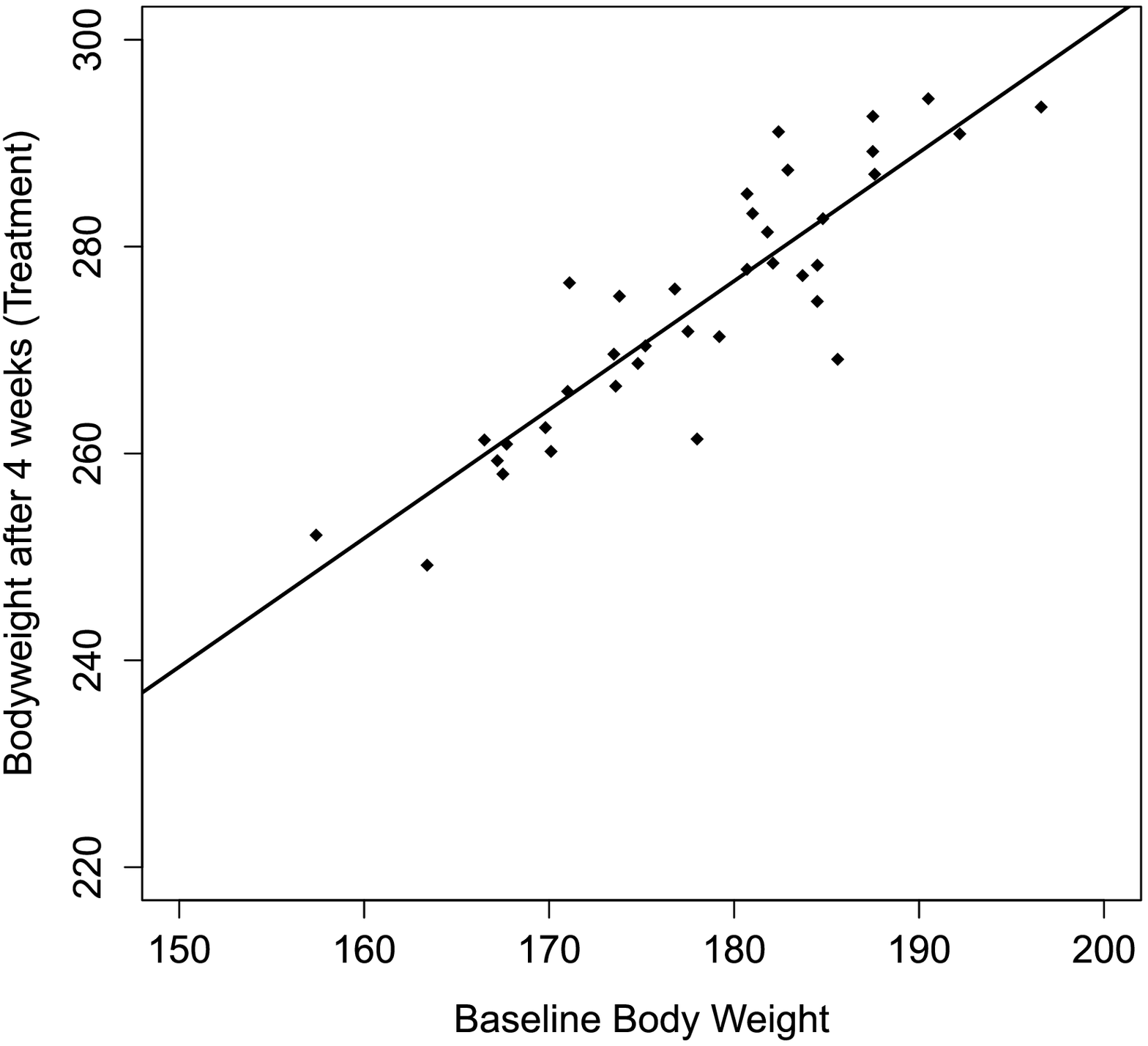} \hfill
\caption{Scatterplots of the bodyweight data of the combined data set (left), vehicle control group (middle) and active treatment group (right).}\label{Fig: scatterplot}
\end{figure}

The boxplots in Figure~\ref{Fig: boxplots} show that the bodyweight distributions at baseline are similar. The bodyweights of the rats after four weeks of treatment seem to be higher under treatment than of those in the vehicle control group. The sample means and the empirical variances of the baseline (M) and response values (Y) are 
\bqa
&\textbf{Baseline} &					\hspace{2cm}   \textbf{After four weeks of treatment}\\ 
\olM_{1\cdot} &=& 177.57, 	\hspace{2cm}  \olY_{1\cdot} = 268.46,\\
\olM_{2\cdot} &=& 176.53, 	\hspace{2cm}  \olY_{2\cdot}  = 271.84,\\
s_{1,M}^2 &=& 127.62, 			\hspace{2cm}   s_1^2 = 183.43,\\
s_{2,M}^2 &=& 119.61, 			\hspace{2cm}   s_2^2 = 258.48.
\eqa 
Thus, based on the empirical variances of the response after four weeks of treatment, assuming equal variances of the data across the two groups is doubtful (183.43 versus 258.48). The baseline values differ slightly in their empirical variances. However, natural variations are normal, even at baseline. Applying the Welch-Satterthwaite $t$-test given in (\ref{ttest}) for testing the null hypothesis $H_0: \mu_1=\mu_2$ yields
\bqa
Baseline: p-value=0.7759,   \hspace{2cm}  \text{After 4 weeks}: p-value = 0.4648
\eqa
 and thus, data do not provide the evidence to reject the null hypotheses at 5\%-level of significance. The $t$-test (assuming equal variances) leads to the same conclusions (baseline p-value = 0.7704; Response p-value = 0.499). We therefore assume that the baseline values are equally distributed across the two groups and that no significant treatment effect exists at 5\% level (after four weeks).  However, the scatterplots of the bodyweights of the rats at baseline and after four weeks of treatment in Figure~\ref{Fig: scatterplot} show that the bodyweights are positively correlated. For illustration, scatterplots of the combined data set (left), vehicle control (middle) and active treatment group are displayed.

Therefore, the $t$-test results diplayed above are doubtful, because the point estimation of $\mu_1$ and $\mu_2$ by their empirical means is biased. 
It can furthermore be seen that the regression coefficients for the combined data set, vehicle control and active treatment groups are similar ($\whp_{Combined}=1.268$, $\whp_{Vehicle} = 0.984$, $\whp_{Treatment} = 1.374$) and therefore the traditional and useful assumption that the regression coefficients are equal across the two groups will be kept for further data evaluations and theoretical investigations.  Of major interest is, however, estimating the adjusted treatment effects $b_1$ and $b_2$ as well as testing the hypothesis that these two effects are identical without assuming that the population variances are equal. In order to gather these information, the data will now be used for the formulation of a general ANCOVA model.

\section{Statistical Model, Hypothesis and Point Estimators}\label{sec:: Model}

\noindent We consider a general two sample ANCOVA model 
\begin{eqnarray}
\bm{Y} = \bm{Xb}+ \bm{Mp} +\bm{\epsilon},
\end{eqnarray}
where 
\begin{eqnarray} \label{Sigma}
E(\bm{\epsilon}) &=& \bm{0}, \quad Var(\bm{\epsilon}) = \bm{\Sigma} =  \bigoplus_{i=1}^2 \sigma_i^2\bm{I}_{n_i} \quad \text{and} \quad E(||\bm{\epsilon}^4||) < \infty.
\end{eqnarray}
Here, $\bm{Y} = (\bm{Y_1}', \bm{Y}_2')'$ denotes the $N\times 1$ response vector of the two samples $\bm{Y}_i = (Y_{i1},\ldots,Y_{in_i})'$ each of size $n_i$, $i=1,2$, $\bm{X} = \bigoplus_{i=1}^2 \mathbf{1}_{n_i}$ denotes the design matrix, $\vb=(b_1,b_2)'$ denotes the vector of fixed treatment effects, $\bm{M}$ is a $N \times L$ matrix collecting the values of the $L$ (fixed) covariates, $\vp=(p_1,\ldots,p_L)'$ denotes the vector of regression coefficients and $\mathbf{1}_{n_i}$ denotes the $n_i \times 1$ vector of 1's, respectively. It is of main interest to test the null hypothesis $H^{\bm{b}}_0: b_1=b_2$ and to compute confidence intervals for $\delta=b_1-b_2$. Furthermore, secondary hypotheses are testing the effects of the $L$ covariates by $ H^{p}_0: p_l=0, l=1,\ldots,L$, seperately. \\

\noindent The parameters $\bm{p}$ and $\bm{b}$ can be estimated using ordinary least squares without bias by
\bqa
\widehat{\bm{p}}&=&\bm{(M'QM)^{-1}M'QY},\\
\widehat{\bm{b}}&=&\bm{(X'X)^{-1}X'(Y-M\widehat{p})},
\eqa
where $\bm{Q} = \bm{{I}_N-X(X'X)^{-1}X'}$ denotes the orthogonal projection onto the column space of $\bm{X}$ see, e.g., \cite{seberlee1977linearregression}. If the covariates in $\bm{M}$ are correlated and thus $\vM$ may not be of full column rank, the inverse $\bm{(M'QM)}^{-1}$ may not exist. However, the linear combination $\delta=b_1-b_2$ is still estimable because $\mathbf{c}'(\bm{\wtX'\wtX})^{-}(\bm{\wtX'\wtX}) = \mathbf{c}'$ holds for any generalized inverse $\bm{(\wtX'\wtX)}^{-}$, where $\bm{\wtX} = (\vX \vdots \vM)$ denotes the partitioned matrix of $\vX$ and $\vM$ and vector $\mathbf{c}=(1,-1,0,\ldots,0)'$. In these cases, the inverse $\bm{(M'QM)}^{-1}$ is replaced by any generalized inverse $\bm{(M'QM)^{-}}$ in the computations above, e.g. by the Moore-Penrose inverse. Finally, the asymptotic distributions of the estimators can be examined. For the ease of representation, define the matrices 
\begin{eqnarray}
\vD &=& (\vX'\vX)^{-1}\vX'-(\vX'\vX)^{-1}\vX'\vM(\vM'\vQ\vM)^{-1}\vM'\vQ  \; \text{and} \label{D}\\
\vA &=& ({\vM'\vQ\vM})^{-1}{\vM'\vQ} \label{A}.
\end{eqnarray}
If the samples are not too unbalanced, i.e. $N\to \infty$ such that $\frac{N}{n_i} \leq N_0 < \infty$, then    

\begin{eqnarray}
&& \quad \sqrt{N} (\widehat{\bm{b}} - \bm{b}) \approx N(\bm{0},N\bm{D}\bm{\Sigma} \bm{D}') \\
 && \quad \sqrt{N} (\widehat{\bm{p}} - \bm{p})\approx N(\bm{0},N\bm{A}\bm{\Sigma} \bm{A}'),
\end{eqnarray}
 where $\vSigma$ is as in (\ref{Sigma}). The covariance matrices ${\bm{\Phi}}=N\bm{A}\bm{\Sigma} \bm{A}'$ and ${\bm{\Psi}}=N\bm{D}\bm{\Sigma} \bm{D}'$, are, however, unknown in practical applications and must be estimated from the data. Their unbiased and consistent estimation is a rather challenging task and will be investigated in detail in the next section. 

\section{Estimation of the variances} \label{sec:: variances}

\noindent The only unknown components of the matrices ${\bm{\Phi}}$ and ${\bm{\Psi}}$ are the variances $\sigma_1^2$ and $\sigma_2^2$ in the setup above. 
In particular, an unbiased and consistent estimator of the variance
\begin{eqnarray}\label{sigma}
{\sigma}_{\vb}^2 = Var(\sqrt{N}(\whb_1 - \whb_2))
\end{eqnarray}
is needed. For the computation of an unbiased estimator, we first compute the detailed structure of ${\sigma}^2_{\vb}$. Let $\vD=(d_{ij})_{i=1,2}^{j=1,\ldots,N}$ be the $2\times N$ generating matrix of $\vwhb$ given in (\ref{D}) and let $A_j = d_{1j}d_{2j}$ for $j=1,\ldots,N$. We obtain with $\vc=(1,-1)'$
\bqan \label{sigmab}
{\sigma}^2_{\vb} &=& Var(\sqrt{N}(\whb_1 - \whb_2)) \nonumber\\
&=& N\vc' \vD\vSigma\vD'\vc \nonumber\\
&=& N \vc' \left(\begin{array}{cc}
\sum\limits_{j=1}^{n_1} d_{1j}^2\sigma_1^2 + \sum\limits_{j=n_1+1}^{N} d_{1j}^2\sigma_2^2 & \sum\limits_{j=1}^{n_1} A_j \sigma_1^2 + \sum\limits_{j=n_1+1}^{N} A_j\sigma_2^2\\
\sum\limits_{j=1}^{n_1} A_j \sigma_1^2 + \sum\limits_{j=n_1+1}^{N} A_j \sigma_2^2 & \sum\limits_{j=1}^{n_1} d_{2j}^2\sigma_1^2 + \sum\limits_{j=n_1+1}^{N} d_{2j}^2\sigma_2^2\\
 \end{array} \right)\vc \nonumber\\
&=& N \left(\sigma_1^2\sum_{j=1}^{n_1} (d_{1j}-d_{2j})^2 + \sigma_2^2\sum_{j=n_1+1}^N (d_{1j}-d_{2j})^2\right)\nonumber\\
&\equiv& N\left( \sigma_1^2 n_1^\ast + \sigma_2^2 n_2^\ast \right).
\eqan
Thus, the variance ${\sigma}_{\vb}^2$ can be expressed as a weighted sum of the variances $\sigma_1^2$ and $\sigma_2^2$. It also follows from the computations above that the estimators $\whb_1$ and $\whb_2$ are highly positively correlated. The correlation among them is implicitly involved in the terms 
\bqan \label{nast}
n_1^\ast = \sum_{j=1}^{n_1} (d_{1j}-d_{2j})^2 \;\;\; \text{and} \; \;\; n_2^\ast=\sum_{j=n_1+1}^N (d_{1j}-d_{2j})^2,
\eqan
which can be interpreted as weighting factors that ensure the consistency of $\whb_1 - \whb_2$ and most importantly, embed their correlations. Furthermore, this result is intriguing and looks familiar when this term is compared with $Var(\olX_{1\cdot} - \olX_{2\cdot}) = \sigma_1^2 /n_1 + \sigma_2^2/n_2$ being used in the Welch $t$-test defined in (\ref{ttest}). \\

\noindent An unbiased estimator of ${\sigma}^2_{\vb}$ is now obtained if unbiased estimators of $\sigma_1^2$ and $\sigma_2^2$ were available. Those can be derived by selecting the corresponding sub-models of model (\ref{model}) and by computing quadratic forms in terms of their residuals. Let $\bm{X}_i = \mathbf{1}_{n_i}$ denote the $n_i \times 1$ vector of 1s and let $\vM_1=\vM_{i=1,\ldots,n_1}$ and 
 $\vM_2=\vM_{i=n_1+1,\ldots,N}$ denote the matrices of the covariates for each group seperately, $i=1,2$. Furthermore, let $\bm{B}_i = (\bm{X}_i\vdots\bm{M}_i)$ denote the two partitioned matrices of $\bm{X}_i$ and the corresponding covariates $\bm{M}_i$, and define the projection matrices
\begin{eqnarray*}
\bm{Q}_i = \bm{I}_{n_i}-\bm{B}_i (\bm{B}_i'\bm{B}_i)^{-1} \bm{B}_i'.
\end{eqnarray*}
Then, unbiased and consistent estimators of the variances $\sigma_i^2$ are given by
\begin{eqnarray} \label{sigmai}
\widehat{\sigma}_i^2 = \bm{Y}_i'\bm{Q}_i\bm{Y}_i/(n_i-1-r(\bm{M}_i)), i=1,2.
\end{eqnarray}
Thus, we obtain an unbiased and consistent estimator of ${\sigma}_{\vb}^2$ given in (\ref{sigmab}) by
\bqan \label{sigmanew}
{\whsigma}^2_{\vb} = N\left( \whsigma_1^2 n_1^\ast + \whsigma_2^2 n_2^\ast \right).
\eqan
These results are summarized below:
\label{thm: unb}
Under the assumptions of model (\ref{model}), the estimators $\whsigma_i^2, i=1,2,$ in (\ref{sigmai}) and ${\whsigma}_{\vb}^2$ defined in (\ref{sigmanew}) are unbiased and $L_2$-consistent, i.e.
\begin{eqnarray}\label{thm: unb}
 E(\widehat{\sigma}_i^2) &=& \sigma_i^2, \;\; \widehat{\sigma}_i^2-{\sigma}_i^2\stackrel{L_2}{\to} 0, n_i \to \infty\; i=1,2,\nonumber\\
  E(\widehat{\sigma}_{\vb}^2) &=& {\sigma}_{\vb}^2, \;\; \widehat{\sigma}_{\vb}^2-{\sigma}_{\vb}^2\stackrel{L_2}{\to} 0, \min\{n_1,n_2\} \to \infty.
\end{eqnarray}
The proof is given in the Appendix. \\

\noindent However, the HCSE estimators of ${\bm{\Phi}}$ and ${\bm{\Psi}}$ are the current state of the art and numerical comparisons of their bias and mean square errors (MSE) are of interest. Numerical and theoretical comparisons will be discussed in the following subsection. 

\subsection{Comparisons with the HCSE variance estimators}
\begin{figure}[h]
	\centering
		\includegraphics[height =  40 ex , keepaspectratio,angle = 0]{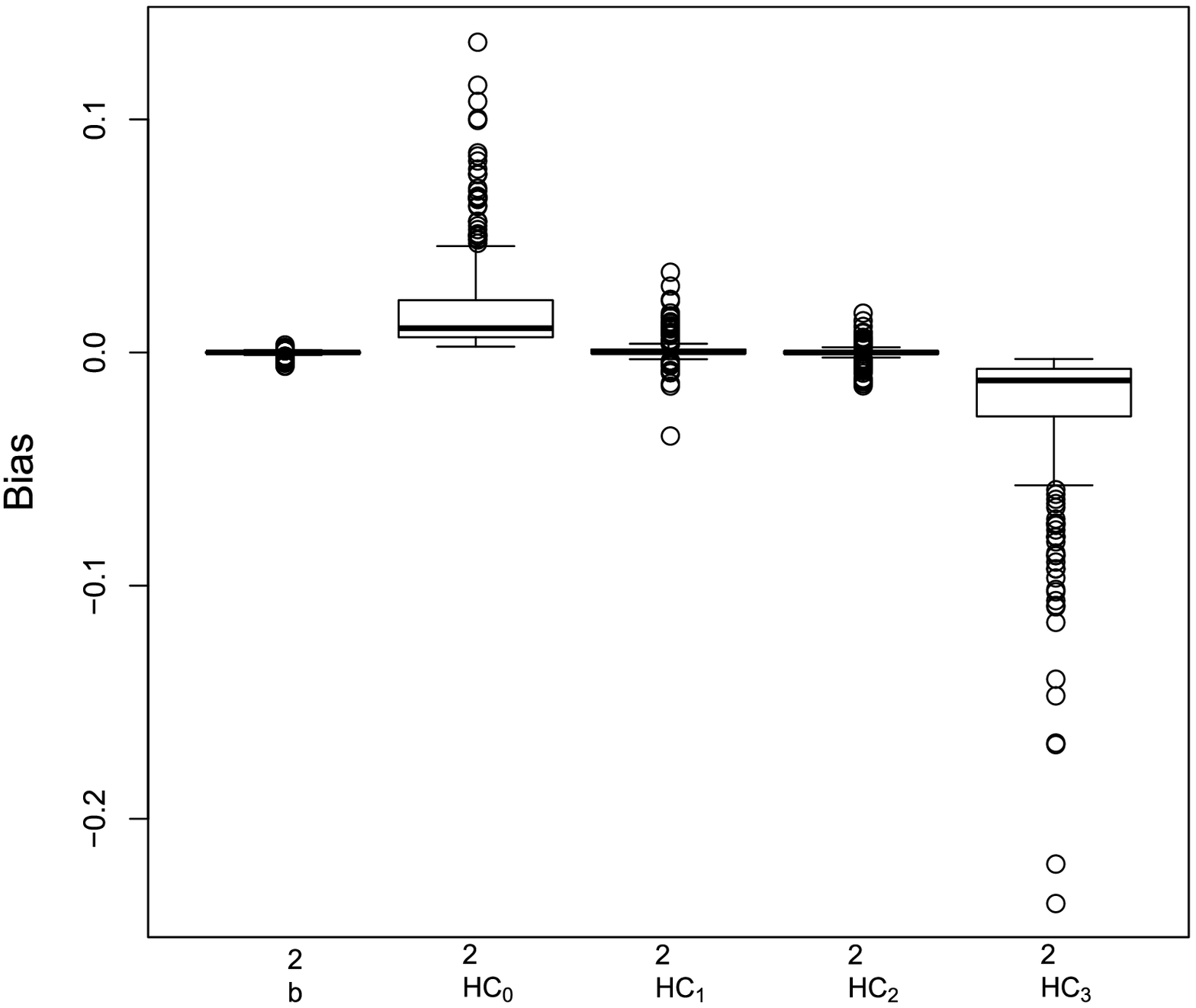}
		\includegraphics[height =  40 ex , keepaspectratio,angle = 0]{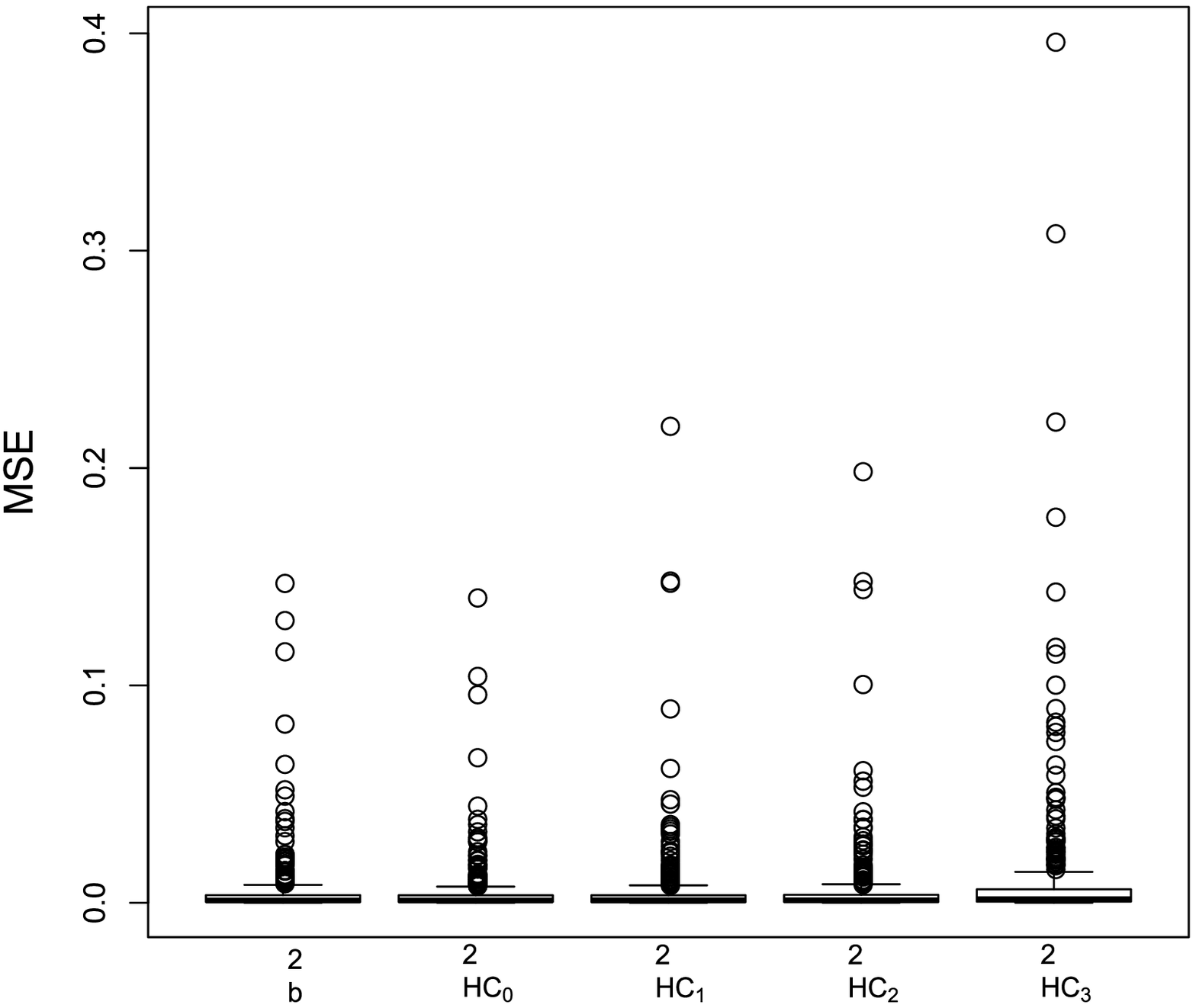}
		\caption{Boxplots of the empirical bias (left) and MSE (right) of $\whsigma^2_{\vb}$ and $\whsigma_{HC_\ell}^2$ given in(\ref{sigmanew})and  (\ref{sigmaHC}), respectively.}\label{Fig: Bias}
\end{figure}

 In order to compare the properties of $\whsigma_{\vb}^2$ with the HCSE estimators, we first re-write the statistical model considered here in the usual HCSE terminology
\begin{eqnarray}
\bm{Y} = \bm{\wtX\beta} + \bm{\epsilon}, \;\text{where}\; \vwtX = (\vX \vdots \vM) \; \text{and} \; \vbeta = (b_1,b_2,p_1,\ldots,p_L)'.
\end{eqnarray}
In this case, the ordinary least squares estimator of $\vbeta$ is given by $\widehat{\bm{\beta}}=(\vwtX'\vwtX)^{-1}\vwtX'\vY$ the covariance matrix of which is 
\bqa
\vGamma=var(\widehat{\bm{\beta}})=(\vwtX'\vwtX)^{-1}\vwtX'{\vSigma}\vwtX(\vwtX'\vwtX)^{-1}.
\eqa
Furthermore, let 
\bqa
\vE_1 &=& \diag\left \{e_1^2,\ldots, e_N^2\right\},\\
\vE_2 &=& \diag\left \{\frac{e_{1}^2}{1-h_{11}},\ldots, \frac{e_{N}^2}{1-h_{NN}}\right \}, \; \text{and}\\
\vE_3 &=& \diag \left\{\frac{e_{1}^2}{(1-h_{11})^2},\ldots, \frac{e_{N}^2}{(1-h_{NN})^2}\right\}
\eqa 
denote the diagonal matrices of the squared and standardized squared residuals, respectively. Here, $h_{ii}$ denotes the $i$th diagonal element of the hat matrix obtained from $\vwtX$. Then, the HCSE estimators as possible candidates for the estimation of $\vGamma$ are 
\bqa
\vwhGamma_{HC_0} &=&(\vwtX'\vwtX)^{-1}\vwtX'\vE_1 \vX(\vwtX'\vwtX)^{-1}, \\
\vwhGamma_{HC_1} &=&\frac{N}{N-L-1}(\vwtX'\vwtX)^{-1}\vwtX'\vE_1\vwtX(\vwtX'\vwtX)^{-1},\\
\vwhGamma_{HC_2} &=& (\vwtX'\vwtX)^{-1}\vwtX'\vE_2\vwtX(\vwtX'\vwtX)^{-1},\\
\vwhGamma_{HC_3} &=& (\vwtX'\vwtX)^{-1}\vwtX'\vE_3\vwtX(\vwtX'\vwtX)^{-1},
\eqa
respectively. More details about the estimators are given in \cite{white1980hetero, chesher1987biashc, furno1996smallsamplehc, cribari2000, bera2002, MacKinnon1985hetero, cribari2000Ferrari, cribari2003Galvao} and references therein. Thus, estimators of ${\sigma}_{\vb}^2$ given in (\ref{sigma}) are given by
\bqan \label{sigmaHC}
{\whsigma}_{HC_\ell}^2 = N \va'\vwhGamma_{HC_\ell}\va, \; \text{where} \; \va=(1,-1,0,\ldots,0)' \; \text{and}\; \ell=0,1,2,3.
\eqan
To investigate the bias of the estimators $\whsigma_{\vb}^2$ and ${\whsigma}_{HC_\ell}^2$ a simulation study has been conducted. Data has been simulated from an independent two-sample ANCOVA model with three covariates and sample sizes $n_1,n_2 \in \{7, \ldots, 40\}$ and variances $\sigma_1^2, \sigma_2^2 \in \{1,3\}$. The bias as well as the MSE of all estimators were computed for each scenario based on 10,000 simulation runs. The results are displayed in the boxplots in Figure \ref{Fig: Bias}. 

It can be seen from Figure~\ref{Fig: Bias} that the estimators $\whsigma_{HC_0}^2$ and $\whsigma_{HC_3}^2$ are substantially biased, especially when sample sizes are small. The bias reduces with increasing sample sizes and depends on variance/sample size allocations. The bias of the estimators $\whsigma_{HC_1}^2$ and $\whsigma_{HC_2}^2$ is way smaller compared to the two others. As expected, the bias of $\whsigma_{\vb}^2$ is about 0. The MSEs of all estimators are very similar and no major differences can be detected. These empirical findings are in concordance with those obtained by \cite{longervin2000}.\\

\noindent Finally, comparing $\whsigma^2_{\vb}$ with $\whsigma_{HC_1}^2$ and $\whsigma_{HC_2}^2$ given in (\ref{sigmaHC}) on a theoretical level, we note that the computation formulas of all these three estimators are similar. We write the quadratic form $\whsigma_i^2$ as a sum of squares of the residuals and obtain
\bqa
\whsigma_i^2 &=& \frac{1}{n_i-1-r(\bm{M}_i)}\bm{Y}_i'\bm{Q}_i\bm{Y}_i \\
&=&  \frac{1}{n_i-1-r(\bm{M}_i)} \vY_i' \bm{Q}_i'\bm{Q}_i\vY_i \\
&=&  \frac{1}{n_i-1-r(\bm{M}_i)} \ve_i'\ve_i \\
&=& \frac{1}{n_i-1-r(\bm{M}_i)} \sum_{k=1}^{n_i} e_{ik}^2.
\eqa
It follows that the normalizing constants used in $\vwhGamma_1$ and $\vwhGamma_2$ are also used in $\whsigma_i^2$, because $n_i-1-r(\bm{M}_i) = \sum_{k=1}^{n_i} (1-h_{kk})$ is the sum of the diagonal elements of the hat matrix of the corresponding sub-model\textemdash since $\vQ_i$is a projection matrix. Thus, $\whsigma^2_{\vb}$ is a bias corrected version of $\whsigma_{HC_1}^2$ and $\whsigma_{HC_2}^2$ in model (\ref{model}). Furthermore, unbiased and consistent estimators of $\bm{\Phi}, \bm{\Psi}$ and $\vGamma$ are given by $\widehat{\bm{\Phi}} = N\bm{A} \widehat{\bm{\Sigma}} \bm{A'} $,  $ \widehat{\bm{\Psi}} = N\bm{D} \widehat{\bm{\Sigma}} \bm{D'}$ and $\vwhGamma = (\vwtX'\vwtX)^{-1}\vwtX'{\vwhSigma}\vwtX(\vwtX'\vwtX)^{-1}$,  where 
\begin{eqnarray}
\widehat{\bm{\Sigma}} = \bigoplus_{i=1}^2 \widehat{\sigma}_i^2\bm{I}_{n_i}.
\end{eqnarray} 

We therefore do not consider the HCSE-based estimators $\whsigma_{HC_\ell}^2$ in further theoretical investigations and data evaluations and will use the unbiased estimator $\whsigma_{\vb}^2$ instead. The point estimators, their asymptotic distributions as well the unbiased and consistent estimation of their parameters can now be used for the derivation of test procedures and confidence intervals. This will be explained in the next section. 

\section{Test Statistics} \label{sec:: tests}
\noindent In this section, different test procedures for testing the two-sided null hypotheses $H_0^{\vb}: b_1=b_2$ as well as $H_0^{\vp}: p_l=0$ for fixed $l=1,\ldots,L,$ will be discussed. In order to test the null hypothesis $H^{\bm{b}}_0:b_1=b_2$, consider the test statistic
\begin{eqnarray}\label{Tnormal} 
T_{\bm{b}} = \sqrt{N} \frac{\whb_1-\whb_2-(b_1-b_2)}{\whsigma_{\vb}} \stackrel{\mathcal{D}}{\to} N(0,1), \; N\to \infty.
\end{eqnarray}
For large sample sizes, the null hypothesis $H^{\bm{b}}_0$ will be rejected at level $\alpha$ of significance, if $|T_{\bm{b}}| \geq z_{1-\alpha/2}$, where $z_{1-\alpha/2}$ denotes the $(1-\alpha/2)$ quantile of the standard normal distribution. An asymptotic $(1-\alpha)$ - confidence interval for $\delta=b_1-b_2$ is given by $CI = \whb_1-\whb_2 \pm \frac{z_{1-\alpha/2}}{\sqrt{N}} \whsigma_{\vb}$. For small sample sizes, however, the test tends to over-reject the null hypothesis. Therefore, we approximate the distribution of $T_{\vb}$ by a central $t_{\kappa}$-distribution and estimate $\kappa$ using Box-type approximation methods.\\

\noindent Note that the estimators $\whsigma_1^2$ and $\whsigma_2^2$ given in (\ref{sigmai}) are independent. Assuming for a moment normally distributed errors, the estimators follow a $\chi^2$-distribution, i.e. $(n_i-1-r(\vM_i)) \whsigma_i^2 \sim \chi_{n_i-1-r(\vM_i)}^2\sigma_i^2$. Hence, it seems to be reasonable to  approximate the distribution of $\whsigma_1^2 n_1^\ast + \whsigma_2^2 n_2^\ast$ by a scaled $\chi_\kappa^2$-distribution, that is $g\cdot \chi_\kappa^2$. The scaling factor $g$ and the degrees of freedom
$\kappa$ are determined in such a way that the expected values and variances of the approximating and the actual sampling distributions coincide. Let $Z\sim \chi_\kappa^2$ and recall that $E(Z) =\kappa, Var(Z)=2\kappa$ and $Var(\whsigma_i^2) = 2\sigma_i^4/(n_i-1-r(\vM_i))$. Therefore, we have to solve the system of linear equations 

\bqa
E\left\{ N\left( \whsigma_1^2 n_1^\ast + \whsigma_2^2 n_2^\ast \right) \right\} &=& N\left\{\sigma_1^2 n_1^\ast + \sigma_2^2 n_2^\ast \right\} \stackrel{!}{=} g\kappa = E(gZ)\\
Var\left\{ N\left( \whsigma_1^2 n_1^\ast + \whsigma_2^2 n_2^\ast \right) \right\} &=& 2 N^2\left\{\frac{\sigma_1^4 n_1^{2,\ast}}{n_1-1-r(\vM_1)} + \frac{\sigma_2^4 n_2^{2,\ast}}{n_2-1-r(\vM_2)} \right\} \\
&\stackrel{!}{=}& 2g^2\kappa = Var(gZ).
\eqa
Replacing the unknown quantities $\sigma_i^2$ in the solution by their empirical counterparts $\whsigma_i^2$, we obtain as estimated degree of freedom 
\bqan \label{kappa}
\kappa = \frac{\left(\whsigma_1^2 n_1^\ast+\whsigma_2^2n_2^\ast\right)^2}{\frac{\whsigma_1^4n_1^{2,\ast}}{n_1-1-r(\vM_1)}+\frac{
\whsigma_2^4n_2^{2,\ast}}{n_2-1-r(\vM_2)}}.
\eqan
It can be readily seen from (\ref{kappa}) that the estimated degree of freedom looks familiar to $\nu$ displayed in (\ref{nu})\textemdash the estimated degree of freedom from the Welch-Satterthwaite $t$-test. Here, the sample variances $s_i^2$ and sample sizes are just replaced by ${\whsigma}_i^2$ and $n_i^\ast$, respectively. Note that $\kappa\to \infty$ if $N\to \infty$ and thus, the approximation procedure is asymptotically correct, even if the normality assumption is violated. For small sample sizes, the distribution of $T_{\vb}$ can be approximated by a central $t_\kappa$-distribution and we reject the null hypothesis $H_0^{\vb}$ at level $\alpha$, if 
\bqan \label{Tkappa}
|T_{\bm{b}}| \geq t_{1-\alpha/2,\kappa},
\eqan 
where $t_{1-\alpha/2, \kappa}$ denotes the $(1-\alpha/2)$-quantile of the central $t_{1-\alpha/2,\kappa}$-distribution with $\kappa$ degrees of freedom. Moreover, approximate $(1-\alpha)$-confidence intervals for $\delta=b_1-b_2$ are given by $CI = \whb_1-\whb_2 \pm \frac{t_{1-\alpha/2, \kappa}}{\sqrt{N}} \whsigma_{\vb}$. The procedure is therefore called \textit{''Welch-Satterthwaite $t$-test with covariates} and denoted as $T_\kappa$ throughout the rest of the paper.

\subsection{Tests for covariate effects and confidence intervals for $p_l$}
Test statistics for testing the secondary null hypotheses $H_0^{\vp}: p_l=0,l=1,\ldots,L,$ can now be derived in a similar way as those for testing $H_0^{\vb}$ discussed in the previous section. First, we compute the variance $Var(\sqrt{N}\whp_l)$ and obtain an unbiased estimator with the same arguments as above. Let $\vA=(a_{ij})_{i=1,\ldots,L}^{j=1,\ldots,N}$ be the $L\times N$ generating matrix of $\vwhp$ given in (\ref{A}) and let $\ve_l$ be the $l$th unit vector. Here, we obtain
\bqa \label{sigmap}
{\sigma}^2_{p_l} &=& Var(\sqrt{N}{\widehat{p}_l}) \\
&=& N\ve' \vA\vSigma\vA'\ve \\
&=& N \left(\sigma_1^2\sum_{j=1}^{n_1} a_{lj}^2 + \sigma_2^2\sum_{j=n_1+1}^N a_{lj}^2\right)\\
&\equiv& N\left( \sigma_1^2 {\wtn}_{1,l} + \sigma_2^2 {\wtn}_{2,l} \right).
\eqa
Hence, the variance of the estimator $\whp_l$ can be written as a weighted sum of the variances $\sigma_1^2$ and $\sigma_2^2$. Replacing these unknown quantities by their unbiased counterparts $\whsigma_1^2$ and $\whsigma_2^2$ yields an unbiased and consistent estimator of ${\sigma}^2_{p_l}$
by
\begin{eqnarray}
{\widehat{\sigma}}^2_{p_l}=N\left( \whsigma_1^2 \widetilde{n}_{1,l} + \whsigma_2^2 \widetilde{n}_{2,l} \right).
\end{eqnarray}
The variance estimator ${\widehat{\sigma}}^2_{p_l}$ can now be used for the derivation of appropriate test statistics for testing $H_0^{\vp}$ and for the computation of confidence intervals for $p_l$, respectively. Consider the test statistic
\bqa
T_{p_l} = \sqrt{N}\frac{\whp_l-p_l}{\sqrt{\widehat{\sigma}^2_{p_l}}},
\eqa
which follows, asymptotically, a standard normal distribution and thus, we reject the null hypothesis $H_0^{\vp}:p_l=0$, if $|T_{p_l}| \geq z_{1-\alpha/2}$. Asymptotic $(1-\alpha)$ - confidence intervals for $p_l$ are given by $CI_l = \widehat{p}_l \pm \frac{z_{1-\alpha/2}} {\sqrt{N}}\widehat{\sigma}_{p_l}$. Simulation studies show, however, that this test tends to over-reject the null hypothesis when sample sizes are rather small. Therefore, we approximate the distribution of $T_{p_l}$ by a $t_{\lambda_l}$-distribution with
\begin{eqnarray} \label{lambda}
\lambda_l = \frac{({\whsigma}_1^2 \widetilde{n}_{1,l}+{\whsigma}_2^2\widetilde{n}_{2,l})^2}{\frac{{\whsigma}_1^4 \widetilde{n}_{1,l}^2}{n_1-1-r({\vM}_1)} + \frac{{\whsigma}_2^4 \widetilde{n}_{2,l}^2}{n_2-1-r({\vM}_2)}}
\end{eqnarray} 
degrees of freedom. Here, $\lambda_l$ is derived in the same way as $\kappa$ in (\ref{kappa}). For small sample sizes, the null hypothesis $H_0^{\vp}$ is rejected at level $\alpha$, if 
\bqan \label{Tlambda}
|T_{p_l}| \geq t_{1-\alpha/2,\lambda_l},
\eqan where $t_{1-\alpha/2,\lambda_l}$ denotes the $(1-\alpha/2)$-quantile of the central $t_{1-\alpha/2,\lambda_l}$-distribution with $\lambda_l$ degrees of freedom. Approximate $(1-\alpha)$-confidence intervals for $p_l$ are given by $CI_l = \whp_l \pm \frac{t_{1-\alpha/2, \lambda_l}}{\sqrt{N}} \whsigma_{p_l}$. \\
Next, the empirical behavior of the developed methods will be investigated in extensive simulation studies.


\section{Simulations} \label{sec: simus}
The test procedures for testing the null hypotheses $H_0^{\vb}$ and $H_0^{\vp}$ developed in the previous section are valid for large sample sizes. Of major interest is investigating their empirical accuracies in terms of controlling the nominal type-1 error rate under the null hypotheses and their powers to detect alternatives when sample sizes are rather small. Extensive simulation studies have been conducted for finding a general conclusion and recommendations for their applicability in practice. All simulations were run using \textit{R} computational environment, version 3.4.0 (www.r-project.org) each with $nsim=10,000$ simulation runs. First, simulation results for $H_0^{\vb}$ will be discussed.
\subsection{Simulation results for $H_0^{b}$} \label{simu: b}

 Recently, \cite{Zimmermann2017Wild} proposed a Wild-Bootstrap test for general factorial ANCOVA designs and their method is also applicable in model (\ref{model}). Since the procedure was shown to be advantageous over White's approach or single wild-bootstrapping in extensive simulations, it will serve as the current state of the art competitor of the Welch-Satterthwaite $t$-test $T_\kappa$  with covariates given in (\ref{Tkappa}). The resampling method is based on the following ideas and will now be briefly explained:
\begin{enumerate}
\item Fix the observed data $\vY$.
\item Randomly generate Rademacher's random signs $W_{ik}$ with $P(W_{ik}=-1) = P(W_{ik}=1)=1/2$.
\item Multiply the residuals with the random signs $W_{ik}$, compute effects $\vwhb$ and $\vwhp$ and $\whsigma_{HC_0}^2$ using the resampling variables.
\item Compute the test statistic (studentized value) from 3.
\item Repeat the above steps a large number of times (e.g. 10K times) and estimate the p-value from the resampling distribution.
\end{enumerate}
For detailed explanations we refer to \cite{Zimmermann2017Wild}. Similar Wild-Bootstrap methods have been used in several inference methods and disciplines, see, e.g., \cite{wu1986jackknife, liu1988wild,mammen1993bootstrap,lin1997non, flachaire2002boot, flachaire2005boot, davidson2008wild, hausman2012inference,mammen2012does, rana2012wild, beyersmann2013weak}.
As additional procedure we considered the classical ANCOVA $t$-test. For the ease of read and graphical presentations, we did not display the simulation results of $T_{\vb}$ using the standard normal approximation as given in (\ref{Tnormal}), because the test is always more liberal than $T_\kappa$, by construction.\\

\begin{figure}[h]
	\centering
		\includegraphics[height =  40 ex , keepaspectratio,angle = 0]{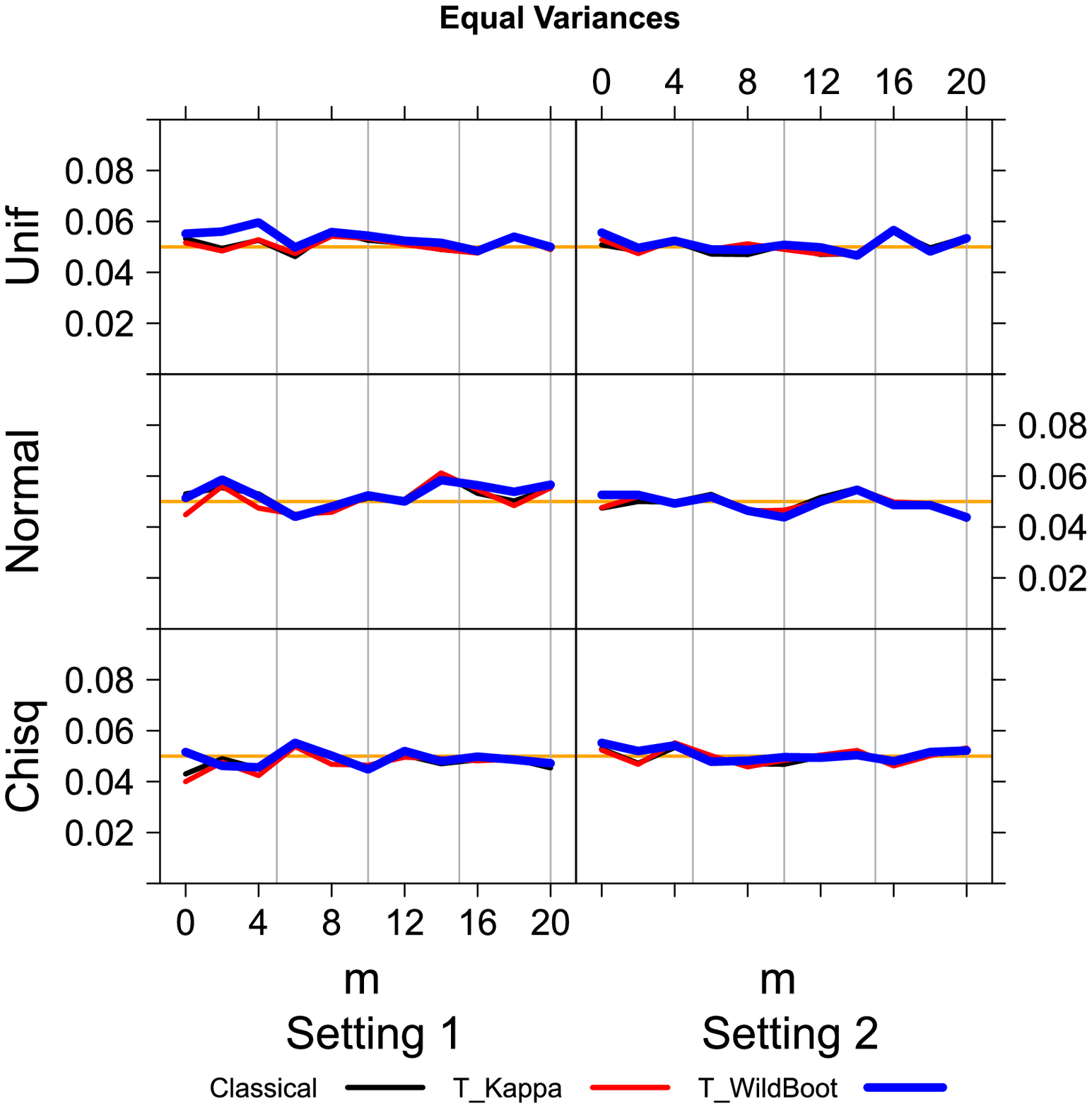}
		\includegraphics[height =  40 ex , keepaspectratio,angle = 0]{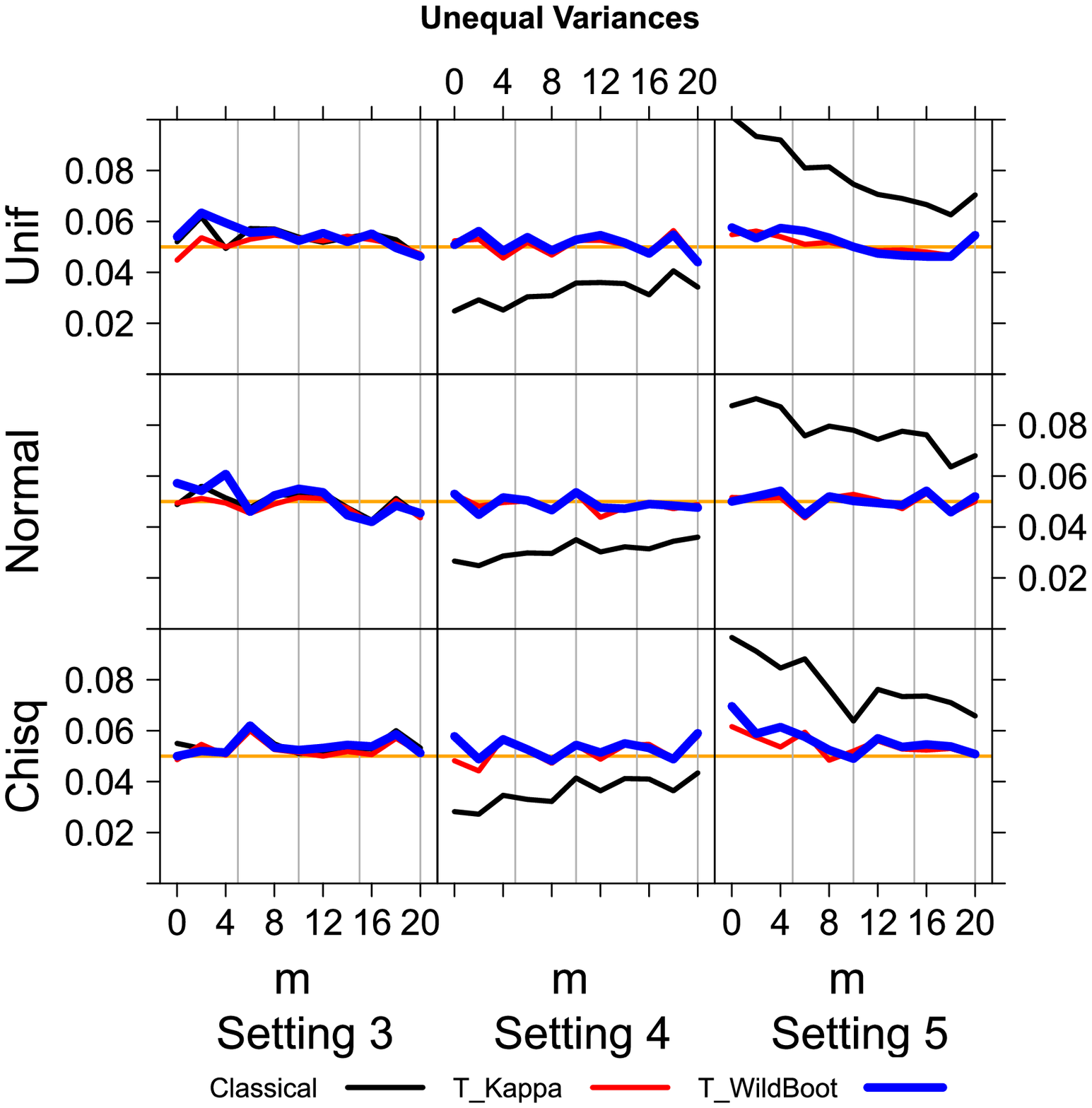}
		\caption{Type-1 error simulation results ($\alpha=5\%$) of the classical ANCOVA $t$-test, $T_\kappa$ defined in (\ref{Tkappa}) and the Wild-Bootstrap method proposed by \cite{Zimmermann2017Wild}.}\label{Fig: type1}
\end{figure}

Data has been generated from
\bqa
\vY= \vX\vb + \vM\vp +\vepsilon,
\eqa
with parameter values $\vb=(10,10)'$, three covariates being the realizations from normal variables with mean $\bm{\mu}=(9,7,5)$ and regression parameters $\vp=(1,0.6,0.7)$. Due to the abundance of different parameter constellations and numbers of covariates included in the model, we keep these settings throughout the simulations and focus on the accuracy of the methods with respect to different error distributions and shapes, small sample sizes, variance heteroscedasticity and unbalanced designs. For the simulation of these scenarios, the error term $\vepsilon$ was generated from standardized normal, uniform and $\chi_7^2$-distributions having variances $\sigma_i^2 \in \{1,3\}$, respectively. We illustrate the performances of these methods when sample sizes increase, i.e., we fix initial sample size allocations of $n_1$ $n_2$ and add an integer $m \in \{0, \ldots,20\}$ for each distributional setting. In total, five different settings will be simulated:
\bqa
\text{\bf Setting 1:} \; (n_1,n_2)&=&(10,10)+m, \;\; (\sigma_1^2, \sigma_2^2) = (1,1), \;\; \text{Balanced, Equal}\\
\text{\bf Setting 2:} \; (n_1,n_2)&=&(10,20)+m, \;\; (\sigma_1^2, \sigma_2^2) = (1,1), \;\; \text{Unbalanced, Equal}\\
\text{\bf Setting 3:} \; (n_1,n_2)&=&(10,10)+m, \;\; (\sigma_1^2, \sigma_2^2) = (1,3), \;\; \text{Balanced, Unequal}\\
\text{\bf Setting 4:} \; (n_1,n_2)&=&(10,20)+m, \;\; (\sigma_1^2, \sigma_2^2) = (1,3), \;\; \text{Unbalanced, Unequal}\\
\text{\bf Setting 5:} \; (n_1,n_2)&=&(20,10)+m, \;\; (\sigma_1^2, \sigma_2^2) = (1,3), \;\; \text{Unbalanced, Unequal}.\\
\eqa 
The nominal type-1 error was set to $\alpha=5\%$ for all simulation runs. The simulation results for all of the scenarios described above are displayed in Figure~\ref{Fig: type1}.

\begin{figure}[h]
	\centering
		\includegraphics[height =  40 ex , keepaspectratio,angle = 0]{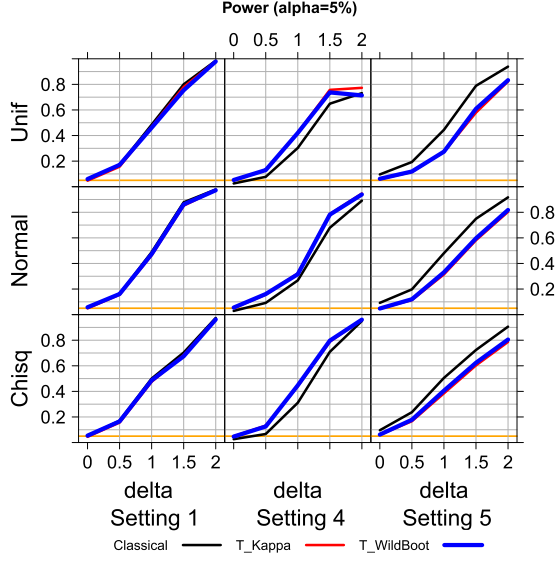}
		\caption{Power simulation results ($\alpha=5\%$) of the classical ANCOVA $t$-test, $T_\kappa$ defined in (\ref{Tkappa}) and the Wild-Bootstrap method proposed by \cite{Zimmermann2017Wild}.}\label{Fig: power1}
\end{figure}

It can be readily seen from Figure~\ref{Fig: type1} that the classical ANCOVA $t$-test controls the size very well when variances across the two groups are equal. This impression changes when the actual variances are different. It tends to be very conservative when the larger sample has the larger variance (Setting 4) and very liberal when variance/sample sizes are negatively allocated, i.e. the larger sample has the smaller variance (Setting 5). This behavior of the test does not improve when sample sizes increase, because the method is based on a pooled variance estimator (which assumes equal variances). It can also be seen that the Welch-Satterthwaite $t$-test $T_\kappa$ controls the nominal type-1 error rate very satisfactorily in all investigated scenarios. The Wild-Bootstrap method proposed by \cite{Zimmermann2017Wild} behaves very similar to the new $t$-test and no major differences in terms of controlling the type-1 error rate can be detected in these selected scenarios. Next, the powers of the methods to detect the alternative $H_1: b_1\not=b_2$ will be investigated.\\
For power investigations, the initial values of the parameter $\vb$ have been shifted by a value $\delta$, i.e.
\bqa
\bm{b}=(10,10+\delta)'\; \; \text{for}\;\; \delta \in \{ 0, 0.5, 1, 1.5, 2 \}
\eqa
in the Settings 1, 4 and 5 described above. For the ease of representation, the sample size increment $m$ was set to 0 for all of these settings. The power curves are displayed in Figure~\ref{Fig: power1}   
and it can be seen that the powers of the new method and the Wild-Bootstrap approach are very similar and almost identical. The conclusion that the classical ANCOVA $t$-test has a higher power than its competitors, however, is incorrect due to its liberality. Based on these empirical findings, we can conclude that the new method is powerful and accurate and even has the same accuracy as the Wild-Bootstrap approach for testing $H_0^{\vb}$. Next, simulation results for testing the hypothesis $H_0^{\vp}$ will be discussed. 
\subsection{Empirical results for $H_0^{p}$}
\begin{figure}[h]
	\centering
		\includegraphics[height =  40 ex , keepaspectratio,angle = 0]{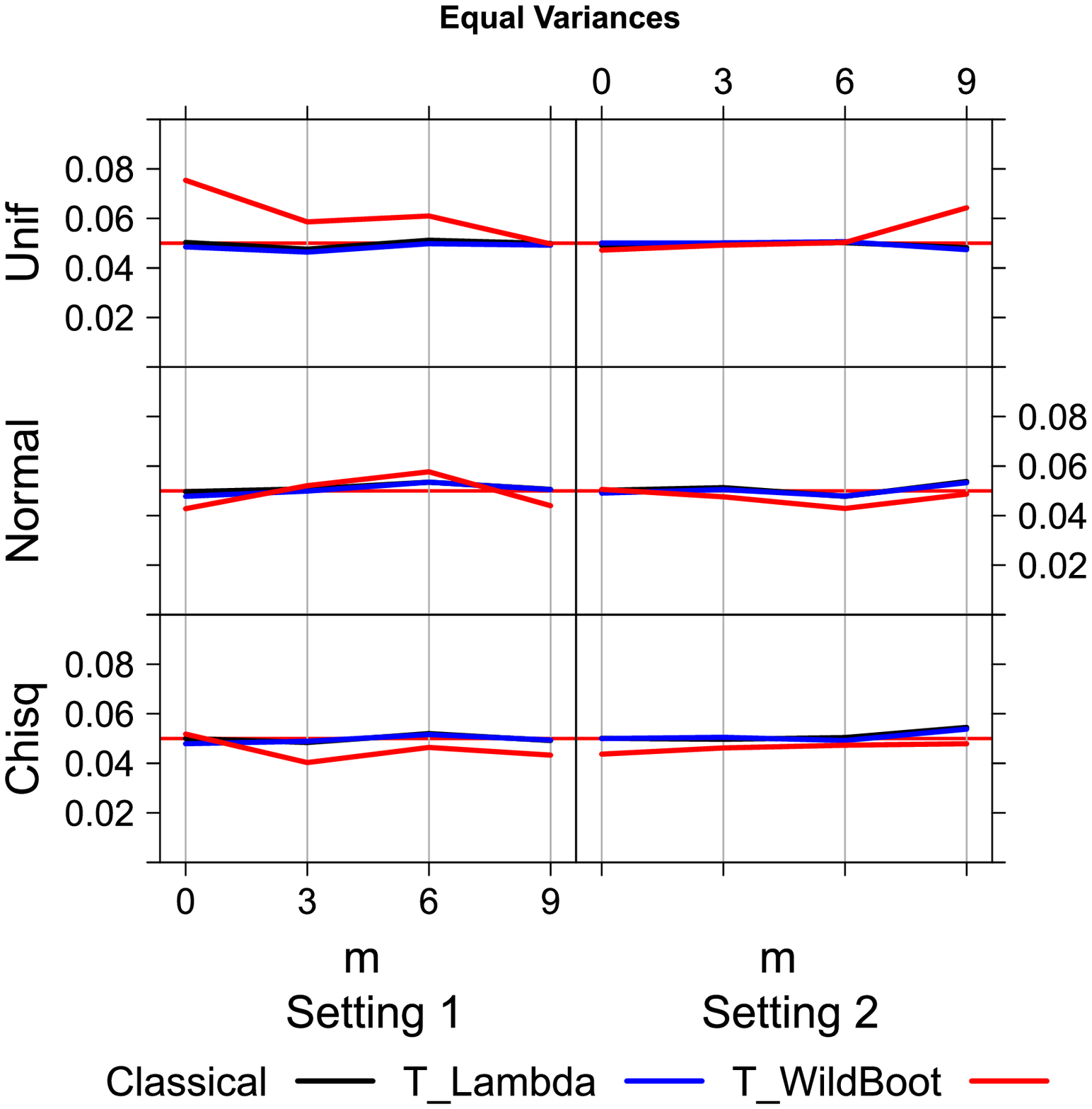}
		\includegraphics[height =  40 ex , keepaspectratio,angle = 0]{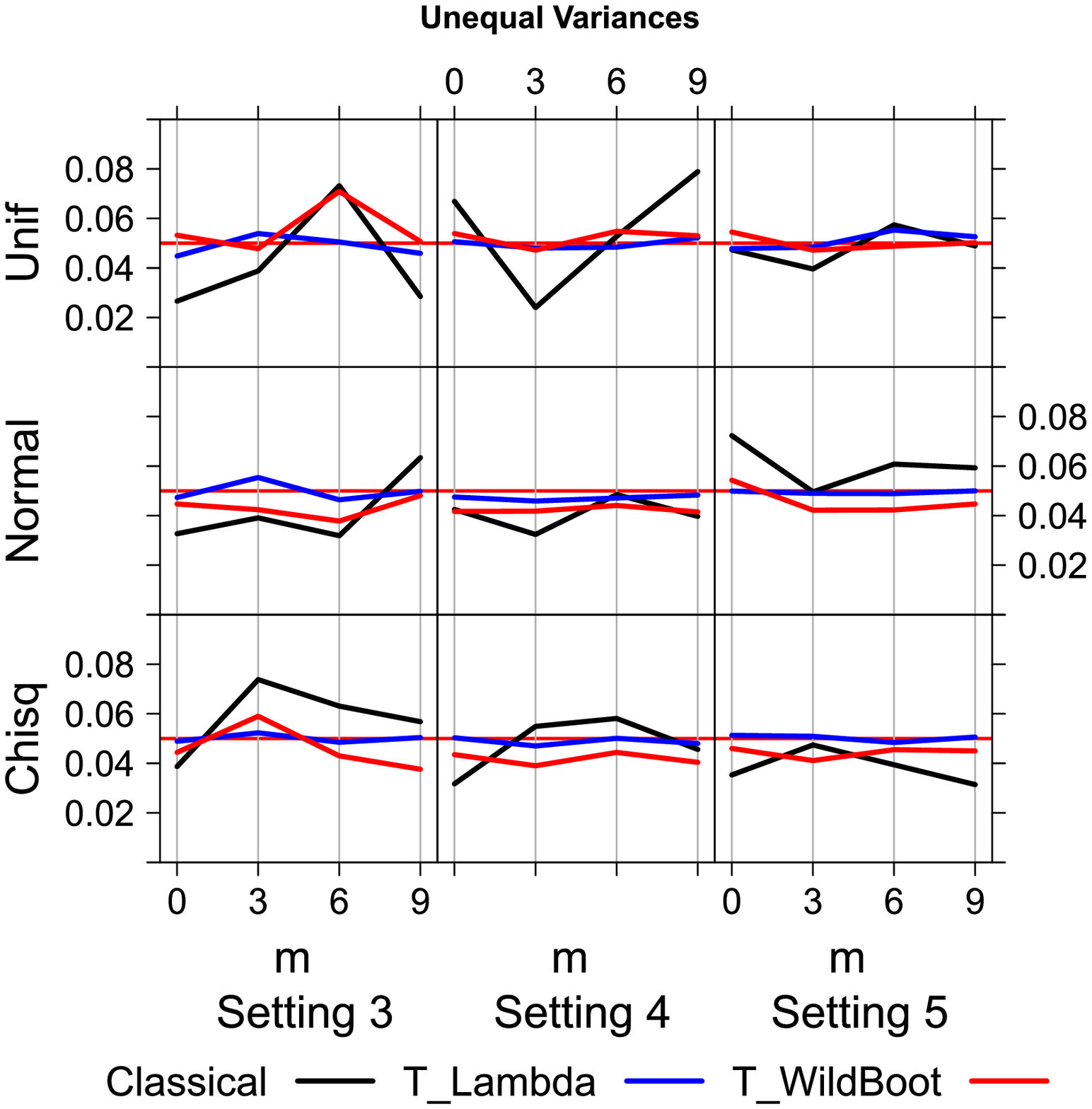}
		\caption{Type-1 error simulation results ($\alpha=5\%$) of the classical ANCOVA $t$-test, $T_\lambda$ defined in (\ref{Tlambda}) and the Wild-Bootstrap method proposed by \cite{Zimmermann2017Wild}.}\label{Fig: type1p}
\end{figure}

In order to test the null hypothesis $H_0^{\vp}: p_l=0$, data has been generated in the same way as described in Section~\ref{simu: b}, with the exception that 
$\vp=(0, 0.6,0.7)'$ was used instead of $\vp=(1,0.6,0.7)'$. Thus, simulation results for $H_0^{\vp}:p_1 = 0$ are reported. We also lowered the sample size increments, because the methods are accurate if $n_1,n_2\geq 20$. Note that \cite{Zimmermann2017Wild} did not investigate inference methods for testing covariate effects in detail. However, their method can be easily modified to that testing problem by using the hypothesis matrix/vector $\vH=(0,0,1,0,0)'$. The simulation results are displayed in Figure~\ref{Fig: type1p}.

\begin{figure}[h]
	\centering
		\includegraphics[height =  40 ex , keepaspectratio,angle = 0]{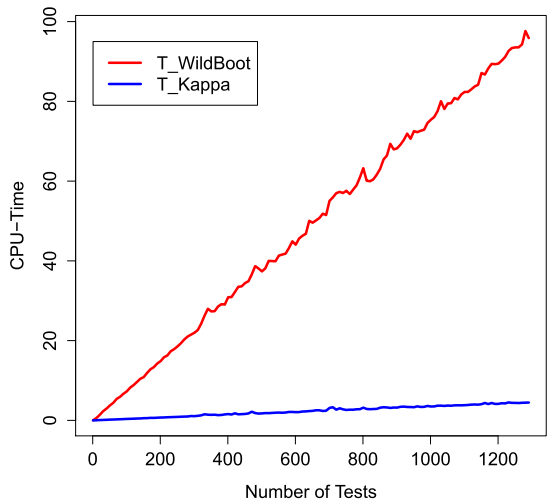}
		\caption{Numerical comparison of the CPU-times for the numerical computation of $T_\kappa$ and the Wild-Bootstrap method for various numbers of tests.}\label{Fig: times}
\end{figure}

It can be readily seen from Figure~\ref{Fig: type1p} that the classical ANCOVA $t$-test controls the nominal type-1 error rate when population variances are equal. This impression changes when the actual variances are different. The classical method does not show a clear tendency towards a liberal or conservative behavior. This occurs, because the method uses the ''classical'' pooled variance estimator
\bqa \label{sigmaC}
\widehat{\sigma}_C^2 = \frac{1}{N-2-r(\vM)}\vY'(\vI_N - \vwtX(\vwtX'\vwtX)^{-1}\vwtX')\vY
\eqa
 for the estimation of $\sigma_{p_l}^2$. In the situations considered here, the expected value of $\widehat{\sigma}_C^2$ is 
\bqa
E(\widehat{\sigma}_C^2) = \frac{1}{N-2-r(\vM)}\left\{\sum_{k=1}^{n_1} (1-h_{kk})\sigma_1^2 + \sum_{k=n_1+1}^{N} (1-h_{kk})\sigma_2^2 \right\}.
\eqa
Thus, the actual bias that is made in the estimation of $Var(\sqrt{N}\whp_l)$ using  $\widehat{\sigma}_C^2$ is
\bqa
&&E(\whsigma_C^2 \wtn_1 + \whsigma_C^2 \wtn_2 - (\sigma_1^2 \wtn_1 + \sigma_2^2 \wtn_2) )\\
 &=& (\wtn_1 + \wtn_2)E(\whsigma_C^2) - (\sigma_1^2 \wtn_1 + \sigma_2^2 \wtn_2)\gtrless 0,
\eqa
depending on the actual values of the covariates, sample sizes and variance allocations. This implies that the variance is either under- or overestimated. Furthermore, the Wild-Bootstrap approach tends to be slightly conservative and shows an ''unstable'' behavior in mostly all of these scenarios. This may occur because only one parameter and its resampling distribution are investigated. Here, the bootstrap distribution may depart from the actual distribution, which results in a liberal behavior of the test\textemdash depending on the actual values of the covariates. On the other hand, the newly developed Welch-Satterthwaite $t$-test controls the nominal type-1 error rate very satisfactorily in all investigated scenarios. Power simulations show that the powers of the competing methods are very similar and the results are therefore omitted. \\
\noindent As a concluding remark, we like to mention that the Wild-Bootstrap method is very numerically intensive which limits its applicability in model selections, screening, multiple comparisons and other big data applications, e.g. in genome wide association studies. As an illustrative example, we display the CPU-times for the numerical computations of $T_\kappa$ and its competitor when several tests are performed in Figure~\ref{Fig: times}. The Wild-Bootstrap approach has been implemented using vectorized programming strategies.

It can be seen from Figure~\ref{Fig: times} that the computation of the Welch-Satterthwaite $t$-test is very fast and increases very slowly for increasing numbers of tests. On the other hand, the computation time of the Wild-Bootstrap method significantly increases with increasing numbers of tests. The same argument also holds in simulation studies and thus, simulating the accuracy of the Wild-Bootstrap in multiple comparison procedures with a large numbers of hypotheses or model selections with large numbers of covariates would be very time consuming and unpractical. 

\section{Data analysis of the example}\label{sec:: evaluation}

The short-term study on bodyweights introduced in Section~\ref{sec::Example} can now be analyzed with the newly developed methods. The point estimators $\whb_1$ and $\whb_2$ of the treatment effects as well as the group specific adjusted variance estimators $\whsigma_1^2$ and $\whsigma_2^2$ are displayed in Table~\ref{tab: estimates}.

\begin{table}[h]
\centering
\caption{Group specific point estimators of the treatment effects and variances of the bodyweights.}\label{tab: estimates}
\begin{tabular}{lccc}\hline
Group & $n_i$ & Treatment Effect $\whb_i$ & Variance $\whsigma_i^2$\\\hline
Vehicle Control & 13 & 41.873 & 65.291\\
Treatment & 39 & 46.576 & 33.392\\\hline
\end{tabular}
\end{table}
The descriptive results displayed in Table~\ref{tab: estimates} are intriguing because (1) even the adjusted variances are different and (2) the impression that the treatment group has a larger variance than the vehicle control group as indicated by the computations in Section~\ref{sec::Example} changes. Here, the variance of the baseline adjusted bodyweights under treatment is way smaller than the adjusted variance in the vehicle control group. This result is intuitively clear by taking a second look at the scatterplots of the data in Figure~\ref{Fig: scatterplot}: A larger amount of variance in the model is explained by the regression in the active treatment group than in the vehicle control group, because data is closer to the regression line and thus, the root mean square error is smaller in the active treatment group. Furthermore, these descriptive results indicate that the assumption of equal variances is doubtful. Next, test statistics, p-values and confidence intervals for testing the hypotheses $H_0^{\vb}$ are displayed in Table~\ref{tab: evaluation}.
\begin{table}[h]
\caption{Effect estimates $\whdelta=\whb_1-\whb_2$, standard errors, test statistics, degrees of freedom (DF), p-values and 95\%-confidence intervals for the bodyweight data.} \label{tab: evaluation}
\begin{tabular}{ccccccr}\hline
Method & Effect & SE & Test Statistic & DF & p-Value & 95\%-CI\\\hline
 $T_\kappa$ & -4.70 &2.43 &-1.94 &14.95 &0.072      &[-9.88;  0.47]\\
 Wild-Boot& -4.70   &2.46 &-1.91  &     -- &0.082   &[-9.81;  0.40]\\
 Classical& -4.70   &2.11 &-2.23 &49 				& 0.031 &[-8.95; -0.46]\\\hline
\end{tabular}
\end{table}
First, it can be readily seen from Table~\ref{tab: evaluation} that the estimated standard errors of the effect $\whdelta=\whb_1-\whb_2$ differ. The classical ANCOVA pooled variance estimator $\whsigma_C^2$ given in (\ref{sigmaC}) (which assumes equal variances), tends to a smaller standard error than the usage of its unbiased competitor $\whsigma_{\vb}^2$ in (\ref{sigmanew}). The HCSE-based estimator as used in the Wild-Bootstrap approach proposed by \cite{Zimmermann2017Wild} is the largest. These differences are reflected in the values of the test statistics and associated p-values: Both the Welch-Satterthwaite $t$-test and the Wild-Bootstrap method provide non-significant results at 5\%-level of significance (p=0.07; p=0.08). The classical ANCOVA $t$-test, however, suggests to reject the null hypothesis. These results are in concordance with the extensive simulation results in Setting 5 (the larger sample has the smaller variance) where a liberal behavior of the classical ANCOVA $t$-test could be seen. The three p-values are, however, close to 5\% and all methods indicate that the bodyweights increase remarkably. Furthermore, as estimated regression effect we obtain $\whp=1.276$. All of the methods reject the null hypothesis $H_0^{\vp}:p=0$. Finally, the empirical group-specific ANCOVA models of the bodyweights can be formulated and are given by
\bqa
Y_{1k} &=& 41.873  + 1.276\cdot M_{1k} + \text{error}(0,65.291),\\
Y_{2k} &=& 46.576  + 1.276\cdot M_{2k} + \text{error}(0,33.392) \;\text{or, in terms of means,}\\
\olY_{1\cdot} &=& 41.872 + 1.276\cdot 177.569,\\
\olY_{2\cdot} &=& 46.576 + 1.276 \cdot 176.533,
\eqa   
which may be useful in model validations and predictions. We note, however, that sample sizes are rather small and a larger trial may be beneficial to justify these results. 
\noindent All of these results indicate, however, that adjusting for covariates is important when those may impact the actual response variables. Applying the $t$-tests without covariates leads to a non-significant result (see Section~\ref{sec::Example}), while the adjusted treatment effects are detected to be significantly different across the two groups.


\section{Discussion}\label{sec:: discussion}
\noindent The Welch-Satterthwaite $t$-test given in (\ref{ttest}) is one of the most prominent and often applied inference method in data evaluations and statistical sciences. The method is known to be somewhat robust and to control the nominal type-1 error rate very well even in unbalanced designs under variance heteroscedasticity when data is roughly symmetrically distributed. In case of skewed distributions, its accuracy depends on the shapes and other distributional characteristics \cite{wilcox2011introduction}. In many experiments, however, covariates may impact the response variables and they may even induce variance heteroscedasticity. Ignoring them may lead to wrong conclusions as could be seen by the illustrative short-term study on bodyweights. Several attempts have been made to generalize the ANCOVA $F$-test or ANCOVA $t$-test, but the situation of variance heteroscedasticity was not considered or the results are not satisfactorily for small sample sizes \cite{quade1967rank, harwell1988empirical, young1995nonpara, wilcox2005approach}. The approaches of \cite{Shields1978investigation, akritas2001ancova, munk2006nonpara, wilcox2016ancova} do not need to assume constant variances between the groups, but they show limits to the number of covariates, i.e, only one or two covariates are permitted in the model. Moreover, their robustness to unbalanced designs is unknown \cite{ananda1998bayesian}. All of these attempts were tempting and motivated us to study general two-samples ANCOVA designs under variance heteroscedasticity. The results are summarized in this paper and entitled as the \textit{Welch-Satterthwaite $t$-test with covariates}, which is a solution for the Behrens-Fisher problem in that specific situation. Here, the numbers of covariates can be arbitrary and they may even be arbitrarily correlated. \\

\noindent The derivation of the method was split in several steps (1) Unbiased estimation of the treatment effects $b_1$, $b_2$ and $\delta=b_1-b_2$ and (2) Unbiased estimation of their standard errors. It turned out that the newly developed variance estimators are a bias-corrected version of the HCSE-estimators and   that the variance of $\whdelta=\whb_1-\whb_2$ can be written as a weighted sum of the variances. This result is surprising, because the estimators are highly positively correlated. The correlation, however, is taken care of by the weights, which are known and linear combinations of the covariates. Thus, the remaining task was the unbiased estimation of the individual variance components. Those were estimated by using independent sub-models. A major advantage of the newly-developed variance estimators is that their sampling distributions can be computed\textemdash at least under normality assumption. Finally, a robust $t$-approximation of the distribution of  the test could be developed. It turned out that the computed degree of freedom is very similar to the well known Satterthwaite degree of freedom. Here, the sample variances and sample sizes are replaced by $\whsigma_i^2$ and the weights $n_{i}^\ast$, respectively. Extensive simulation studies show that the new method is as accurate and powerful as the recently proposed Wild-Bootstrap version by \cite{Zimmermann2017Wild}. It also turned out that the Welch-Satterthwaite $t$-test with covariates tends to be slightly more accurate than the Wild-Bootstrap version when the impact of the covariates is tested. Overall, the new method is numerically fast, feasible to compute and the computational formulas are available in a closed form. This is a major advantage of the new method compared to the Bootstrap version. \\

\noindent Comparing the Wild-Bootstrap test and the new method from an educational point of view, it is worth to mention that the new test could be used in introductory classes in Statistics, linear model theory and in other teaching purposes. The theoretical results developed in this paper are of interest of their own. In particular, the illustrative data example clearly shows that adjusting for covariates is important in statistical practice.  

\noindent Throughout the paper we assumed that the groups have identical slope parameters, that is, effect sizes do not depend on the values of the covariates. Note that the model can be generalized to group-specific slope parameters by considering the model
\bqa
\vY = \vX\vb +\left(\bigoplus_{i=1}^2\vM_i\right) \vwtp +\vepsilon, \; \text{where}\; \vwtp=(p_{11},\ldots,p_{1L},p_{21},\ldots,p_{2L})'.
\eqa
Unbiased estimators of the variance components are now obtained by modifying the matrices $\vD$ and $\vA$ defined in (\ref{D}) and (\ref{A}) accordingly. All of the methods considered in the paper are mean-based, i.e., an accurate behavior of the methods when data follow are very skewed distribution cannot be expected. Robust methods that do not require identical slope parameters and simultaneously allow heteroscedasticity have recently proposed by \cite{wilcox2011introduction}. General robust estimation approaches are also discussed in \cite{hampel2011robust, huber2011robust, staudte2011robust}.   

\noindent In the present paper we assumed that the covariates are fixed. Developing unbiased variance estimators in case of random covariates as well as generalizations to completely variance heteroscedastic designs will be part of future research.

\section*{Appendix. Proofs}
\subsection*{A.1. Proof of (\ref{thm: unb})}
\noindent Let $\bm{B}_i (\bm{B}_i'\bm{B}_i)^{-1} \bm{B}_i'=\bm{P}_{n_i}$, $i=1, 2$ denote the projection matrix for each group in the linear model separately. Computing the expectation of the quadratic form yields
\bqa
&&(n_i-1-r(\bm{M}_i))E(\widehat{\sigma}_i^2) =E(\bm{Y}_i'\bm{Q}_i\bm{Y}_i) \\
&=&E(\bm{Y}_i'(\bm{I}_{n_i}-\bm{P}_{n_i})\bm{Y}_i) \\
&=&(\bm{Xb}+ \bm{Mp})'(\bm{I}_{n_i}-\bm{P}_{n_i})(\bm{Xb}+ \bm{Mp})+tr((\bm{I}_{n_i}-\bm{P}_{n_i})\sigma_i^2\bm{I}) \\
&=&(\bm{Xb}+ \bm{Mp})'((\bm{Xb}+ \bm{Mp})-(\bm{Xb}+ \bm{Mp}))+(n_i-1-r(\bm{M}_i))\sigma_i^2   \\
&=&(n_i-1-r(\bm{M}_i))\sigma_i^2,\quad i=1, 2. 
\eqa

\noindent Thus,  $\widehat{\sigma}_i^2$ is an unbiased estimator of $\sigma_i^2$, $i=1, 2$. Next, the consistency of the variance estimators will be shown. We compute the variance of the quadratic form $\whsigma_i^2$ and obtain   

\bqa
Var(\bm{Y}_i'\bm{Q}_i\bm{Y}_i)=(\mu_4-3\sigma_i^4)\bm{q}_i'\bm{q}_i+2\sigma_i^4tr(\bm{\bm{Q}_i}^2)+4\sigma_i^2\bm{\mu}_i'{\bm{Q}_i}^2\bm{\mu}_i+4\mu_3\bm{\mu}_i'\bm{Q}_i\bm{q}_i, 
\eqa
where $\bm{\mu}_i=E(\bm{Y}_i)$, $\bm{q}_i=diag\{\bm{Q}_i\}$, the vector of diagonal elements of $\bm{Q}_i$. Here, $\mu_3$ and $\mu_4$ denote the skewness and kurtosis of the error distributions, respectively. Using the properties of projection matrix $\bm{P}_{n_i}$, we get $0\leq\bm{q}_i'\bm{q}_i\leq n_i+tr(\bm{P}_{n_i})=n_i+1+r(\bm{M}_i)$ and $tr(\bm{Q}_i^2)=tr(\bm{Q}_i)=n_i-1-rank(\bm{M}_i)$. Since ${\bm{Q}_i}^2=\bm{Q}_i, {\bm{Q}_i}^2\bm{\mu}_i=\bm{Q}_i\bm{\mu}_i=(\bm{I}_{n_i}-\bm{P}_{n_i})\bm{\mu}_i=0$. Furthermore, $\bm{\mu}_i'\bm{Q}_i\bm{q}_i=(\bm{Q}_i\bm{\mu}_i)'\bm{q}_i=0$. In conclusion, the $L_2$-convergence follows, because 
\bqa
&& Var(\bm{Y}_i'\bm{Q}_i\bm{Y}_i)/(n_i-1-rank(\bm{M}_i))^2\xrightarrow{L_2}0 , n_i \to \infty.
\eqa

\begin{table}[h]
\centering
\caption{Bodyweight data (Vehicle Control) of the short-term bodyweight study. }\label{dat: bodyweight0}
\begin{tabular}{cccc}
Animal & Dose & Baseline & Week 4 \\
  1 &   0 & 174.20 & 261.00 \\ 
    2 &   0 & 184.20 & 282.90 \\ 
    3 &   0 & 176.90 & 269.80 \\ 
    4 &   0 & 177.00 & 260.80 \\ 
    5 &   0 & 177.10 & 266.30 \\ 
    6 &   0 & 166.90 & 256.10 \\ 
    7 &   0 & 163.90 & 249.50 \\ 
    8 &   0 & 187.60 & 290.50 \\ 
    9 &   0 & 157.40 & 263.50 \\ 
   10 &   0 & 177.30 & 256.30 \\ 
   11 &   0 & 196.00 & 289.30 \\ 
   12 &   0 & 174.50 & 261.00 \\ 
   13 &   0 & 195.40 & 283.00 \\ 
	\end{tabular}
	\end{table}
\newpage
\begin{table}[h]
\centering
\caption{Bodyweight data (Treatment) of the short-term bodyweight study. }\label{dat: bodyweight1}
\begin{tabular}{cccc}
Animal & Dose & Baseline & Week 4 \\
 14 &   1 & 171.00 & 266.00 \\ 
   15 &   1 & 185.60 & 269.10 \\ 
   16 &   1 & 187.50 & 292.60 \\ 
   17 &   1 & 176.80 & 275.90 \\ 
   18 &   1 & 175.20 & 270.40 \\ 
   19 &   1 & 182.90 & 287.40 \\ 
   20 &   1 & 173.80 & 275.20 \\ 
   21 &   1 & 181.80 & 281.40 \\ 
   22 &   1 & 184.50 & 274.70 \\ 
   23 &   1 & 181.00 & 283.20 \\ 
   24 &   1 & 167.20 & 259.30 \\ 
   25 &   1 & 190.50 & 294.30 \\ 
   26 &   1 & 170.10 & 260.20 \\ 
   27 &   1 & 196.60 & 293.50 \\ 
   28 &   1 & 192.20 & 290.90 \\ 
   29 &   1 & 180.70 & 285.10 \\ 
   30 &   1 & 183.70 & 277.20 \\ 
   31 &   1 & 182.40 & 291.10 \\ 
   32 &   1 & 167.50 & 258.00 \\ 
   33 &   1 & 180.70 & 277.80 \\ 
   34 &   1 & 179.20 & 271.30 \\ 
   35 &   1 & 163.40 & 249.20 \\ 
   36 &   1 & 184.50 & 278.20 \\ 
   37 &   1 & 167.70 & 260.90 \\ 
   38 &   1 & 173.60 & 266.50 \\ 
   39 &   1 & 166.50 & 261.30 \\ 
   40 &   1 & 184.80 & 282.70 \\ 
   41 &   1 & 187.60 & 287.00 \\ 
   42 &   1 & 182.10 & 278.40 \\ 
   43 &   1 & 169.80 & 262.50 \\ 
   44 &   1 & 171.10 & 276.50 \\ 
   45 &   1 & 187.50 & 289.20 \\ 
   46 &   1 & 157.40 & 252.10 \\ 
   47 &   1 & 178.00 & 261.40 \\ 
   48 &   1 & 177.50 & 271.80 \\ 
   49 &   1 & 149.30 & 229.00 \\ 
   50 &   1 & 174.80 & 268.70 \\ 
   51 &   1 & 173.50 & 269.60 \\ 
   52 &   1 & 144.80 & 222.30 \\ \hline
		\end{tabular}
	\end{table}

\newpage

\bibliography{Citations}

\end{document}